\documentclass[preprint,10.5pt,authoryear]{elsarticle}
\usepackage[perpage]{footmisc}
\usepackage{placeins}
\usepackage{float} 
\usepackage{setspace} % 用于设置行距  
\usepackage{multirow} % 放在导言区
\usepackage{array} % 用于更多表格功能
\usepackage[a4paper, top=2cm, bottom=2cm, left=1.6cm, right=1.6cm]{geometry}
\usepackage{graphicx} % For \resizebox
\usepackage{booktabs}
\usepackage{diagbox}
\usepackage{caption}
\usepackage{hyperref}  % 用于生成可点击的链接
\usepackage{xcolor}    % 用于设置颜色
\definecolor{blue}{RGB}{33, 120, 250}  % 深蓝色
\definecolor{purple}{RGB}{148, 0, 211} % 定义紫色
\newcommand{\rev}[1]{\textcolor{black}{#1}} % 如果保持原来黑色，则取消注释
\hypersetup{
    colorlinks=true,    % 激活颜色链接
    linkcolor=deepblue, % 内部链接的颜色（如图表引用）
    filecolor=magenta,  % 本地文件链接的颜色
    urlcolor=cyan,      % URL 链接的颜色
}
\usepackage{amssymb}
\usepackage{amsmath}
\journal{}
\makeatletter
\def\ps@pprintTitle{%
  \let\@oddhead\@empty
  \let\@evenhead\@empty
  \def\@oddfoot{\reset@font\hfil\thepage\hfil}%
  \let\@evenfoot\@oddfoot}
\makeatother

\usepackage{fancyhdr}
\begin{document}
\begin{frontmatter}



\title{Collaborative governance of cyber violence: A two-phase, multi-scenario four-party evolutionary game and SB$\mathrm{I}_1$$\mathrm{I}_2$R public opinion dissemination} %% Article title

\author{Xiaoting Yang\fnref{label1}}
\ead{15625085312@163.com}
%%\fntext[label2]{}
\address[label1]{{School of Safety Science and Emergency Management, Wuhan University of Technology, Wuhan, China }}
            
\author{Wei Lv\corref{cor1}\fnref{label1}}
\ead{weil@whut.edu.cn}
%%\fntext[label2]{}
 \cortext[cor1]{corresponding author}

\author{Ting Yang\fnref{label1}}
\ead{19503853765@163.com}  

\author{Bart Baesens\fnref{label3}}
\ead{bart.baesens@kuleuven.be}            
\address[label3]{{Faculty of Economics and Business, KU Leuven, Leuven, Belgium}}

\begin{abstract}
Cyber violence severely disrupts public order in both cyberspace and the real world. Existing studies have gradually advocated collaborative governance but rely on macro-level theoretical analyses. This study integrates micro- and macro-level perspectives to propose a two-stage, multi-scenario governance mechanism for cyber violence. In the first phase, a multi-scenario evolutionary game model with four parties involved in cyber violence was developed based on evolutionary game theory. Matlab simulations show that under strong government regulation, moderate levels of punishment implemented by the government against the online media that adopt misguidance strategies can achieve the most desirable stable state. In the second phase, the role of bystanders was introduced by integrating communication dynamics theory, and emotional factors were considered alongside game strategies. This led to the development of a new $SB{{I}_{1}}{{I}_{2}}R$ model for public opinion dissemination in cyber violence. Netlogo simulations found that increasing the “correct guidance” strategy by the online media reduces the influence of cyber violence supporters and the time it takes for their nodes to drop to zero, but does not significantly shorten the time for the peak to occur. Comparatively, collaborative intervention between the online media and the government was most effective in curbing public opinion, followed by the government’s independent “strong regulation.” Relying solely on the online media’s “correct guidance” produced the weakest effect. Finally, this mechanism was applied to a case study, and a multi-stage, multi-scenario analysis based on life cycle theory enhanced its practical applicability.

This is the accepted manuscript version of the article. The final published version is available at \url{https://doi.org/10.1016/j.ipm.2025.104242}.

\end{abstract}

%% Keywords

\end{frontmatter}

\section{Introduction}

\label{sec1}
%% Labels are used to cross-reference an item using \ref command.
In the 1990s, when the Internet was just emerging, early signs of cyber violence were already appearing in countries such as the United Kingdom and the United States. Cyber violence has become a governance challenge faced globally. According to official reports from the International Telecommunication Union, as of 2023, the global number of Internet users is approximately 5.4 billion people, accounting for about 67\% of the world’s population \citep{International}. The widespread adoption of the Internet has lowered the cost of Internet use for the public, enabling users to anonymously attack others anytime and anywhere \citep{zhao_modeling_2022}. This is one of the key differences between cyber violence and traditional forms of violence.\rev{Cyber violence is regarded as an extension of social violence in cyberspace. Interest groups behind online violence exacerbate social conflicts through improper means, threatening national ideological security, cybersecurity, and social security.}

Violent behavior has shifted from offline to online spaces, and the perpetrator has broken through the limits of physical isolation, further violating victims' rights to privacy, reputation, and property \citep{niki_cyber-attacks_2022}, and even triggering serious mental breakdowns or suicides.\rev{The traditional governance model for cyber violence heavily relies on unilateral government regulation, such as improving the accountability system through legislation \citep{suhartono_cyber_2021}. However, this command-control governance pattern excessively focuses on the binary opposition between the government and perpetrator, making it difficult to explain the interest games and governance effectiveness among multiple actors \citep{Weissman}. Existing evolutionary game theory studies mostly concentrate on dyadic models between the government and perpetrator or between the victim and perpetrator. The evolution of cyber violence is essentially a strategic game among multiple actors driven by interest demands. \citet{Weissman} pointed out that the victim is the directly affected party, and they weigh their response strategies between rights infringement and the cost of seeking rights. The perpetrator is the agent who carries out the behavior, seeking to maximize attack benefits through anonymity and technical tools \citep{mahmud2023}. \citet{mccombs2014new}’s agenda-setting theory indicates that the priority and framing of issues in online media reports directly affect the distribution of public attention, thus influencing victims’ defense strategies and perpetrators’ risk expectations. The online media, as information disseminators, choose reporting tendencies between traffic competition and social responsibility \citep{christner2025}. The government, as the rule-maker, needs to balance freedom of expression and the maintenance of order \citep{tao2025}. Thus, this study constructs a four-party game model involving the victim, the perpetrator, the online media, and the government. Analyzing the strategic interactions among these agents helps to better understand the dynamic evolution of cyber violence.}

\rev{One of the key factors in the governance of cyber violence is the government’s punishment mechanism, especially how to regulate the perpetrator and online media through legislative policies.} Recently, scholars have focused on collaborative governance patterns \citep{rodriguez-rodriguez_how_2020,chin_comparative_2022}, but most studies analyze governance strategies linearly and statically \citep{Zhu2024}, neglecting subjects' limited rationality  and dynamic interactions in the governance process. Existing research reveals the significant impact of government punishment strategies on the behavior of other entities\rev{but overlooks the dynamic relationship between punishment targets and severity, limiting policy effectiveness}. To address this gap, this study constructs a game model based on the advantages of evolutionary game theory in group behavior analysis, incorporating both the target and severity of government punishment within the same framework. It systematically analyzes the dynamic impact of these two factors on the strategy choices of different entities, offering theoretical support for cyber violence governance.

\rev{In fact, the game process in cyber violence is not only reflected in the behavioral choices of multiple actors but also affects the diffusion of public opinion. The game strategies of the government and online media, in particular, partly determine the scope and direction of opinion spread, such as influencing the diffusion threshold and public emotions and responses. In turn, changes in public opinion on cyber violence also affect the strategies of agents. This influence may be short-term and direct, or long-term and indirect. The game process among actors and the diffusion of public opinion are closely related in cyber violence diffusion. The game process reveals the internal logic of multi-party interest conflicts, while public opinion diffusion is the external expression of these conflicts amplified by media. Their interaction jointly drives either the escalation or mitigation of violent events. The diffusion of cyber violence is closely linked to the complex opinion environment. A deep analysis of its generation mechanism and evolution is essential for designing effective governance and intervention strategies.} Recently, the role of bystanders in cyber violence has received widespread attention, particularly regarding their influence on public opinion propagation and the escalation of online conflicts \citep{liu2021}. A common research trend is the exploration of bystander intervention behaviors, which are crucial for mitigating the spread of cyber violence and improving governance measures \citep{liu2021,zhao_cyberbullying_2023}.
Bystanders can either prevent the spread of violence through intervention or exacerbate it through silence or complicity. However, existing research often analyzes bystander behavior statically, overlooking the dynamics and heterogeneity of their attitudes. For example, \citet{zhao_cyberbullying_2023} explored bystander reactions to cyber violence incidents in a specific social media context. This approach controlled variables but weakened the influence of real-life scenarios and individual emotions on behavior. In fact, bystander behavior is often influenced by multiple factors, including personal emotions, group pressure, media perspectives, and intervention policies \citep{yan_stochastic_2021}. Infectious disease dynamics theory, due to its applicability in information diffusion, has been widely applied to the public opinion propagation. For example, \citet{poiitis_aggression_2021} constructed the Linear Threshold (LT) diffusion model to simulate opinion propagation in the context of cyber violence, but it overlooks specific scenarios and emotional factors. \citet{arato_risk_2022} pointed out that the social dynamic emotional processes of cyber violence still require further research. This study innovatively incorporates bystanders into the diffusion dynamics model, considering their emotional interactions. It proposes an optimized dynamic model of cyber violence public opinion propagation to reveal its formation, diffusion, and decline, while exploring the evolution of netizens' opinions under different intervention strategies by the online media and government. To the best of our knowledge, this is a rare attempt to consider detailed intervention strategies.

\rev{However, there is a disconnection between macro and micro analyses in current cyber violence research, leading to a one-sided understanding of the phenomenon.} Macro studies focus on social structures, policies, and legal frameworks \citep{amro_integrated_2023}\rev {but it failed to explain the behavioral roots and the mechanisms of opinion diffusion.} Micro studies mostly examine personal psychology and social networks \citep{huang_what_2023}\rev{but tend to ignore macro factors. In fact, macro and micro factors are both closely related to actors’ strategic choices and the evolution of public opinion in cyber violence. Macro factors provide the rules and constraints that shape the boundaries and direction of strategic interactions. Policy orientation also influences the trajectory of public opinion. Micro factors determine the specific choices made by actors and affect the spread of opinion.} Recent studies have explored frameworks combining macro and micro factors \citep{myers_cyberbullying_2019-1}. For instance, \citet{Patel} emphasized the role of individual traits, attitudes, and social contexts in preventing and intervening in cyber violence.

\subsection{Research questions}
Based on the above findings, several pressing and intriguing questions naturally arise:  

RQ1: What are the strategic interactions among\rev{the victim, perpetrator, online media, and government} within the cyber violence ecosystem, and how do they influence each other? 

RQ2: How does the government’s punitive mechanism function within the cyber violence ecosystem, and how does it affect the behavioral choices of other actors? 

RQ3: How do different intervention scenarios impact the evolution of cyber violence public opinion and the effectiveness of governance measures? 
\subsection{Research objectives}
These questions lead to the primary objectives of this study. This study proposes a collaborative governance framework for cyber violence that integrates both macro- and micro-level perspectives, incorporating multiple factors such as government policies, social influences, individual emotions, and behaviors. The framework consists of a game-theoretic phase and a public opinion propagation phase, offering new insights into the construction of a more coordinated cyber violence governance mechanism. Specifically, the first objective is to examine the strategic interactions and mechanisms among the victim, perpetrator, government, and online media within the cyber violence ecosystem. Analyzing the game relationships among these stakeholders enables a more comprehensive understanding of the evolutionary game dynamics of cyber violence, providing theoretical support for the formulation of effective governance strategies. The second objective is to analyze the impact of the government’s punitive mechanism on the cyber violence game system. Although relevant administrative sanctions have been widely implemented, few studies have systematically examined the differentiated effects of government punishments on the online media and perpetrator. The third objective is to explore the influence of the government and online media interventions on the evolution of cyber violence public opinion. By coupling micro-level behavioral mechanisms with macro-level governance effectiveness, this study aims to provide policymakers with scientifically grounded insights for decision-making.

The remainder of this paper is organized as follows: Section 2 provides an overview of past research. Section 3 describes the research design process for the long-term mechanisms of cyber violence collaborative governance, focusing on the construction and analysis of the four-party evolutionary game and public opinion dissemination models. Section 4 details simulation analyses of the four-party evolutionary game and public opinion dissemination phases. In Section 5, the proposed cyber violence collaborative governance mechanism is applied to a case study of cyber violence. The discussions and conclusions are presented in section 6 and 7, respectively.

\section{Literature review }

\subsection{Definition of cyber violence }
\label{subsec1}
Cyber violence is characterized by its scale, aggregation, harmfulness, and aggressiveness \citep{paat_psycho-emotional_2020}, and is a social norm deviation phenomenon resulting from multiple factors, including societal, technological, media, and user-related causes. At the societal level, cyber violence reflects the distortion of social values and moral degradation. On the technological level, the anonymity, immediacy, and ubiquity of the Internet provide a breeding ground for cyber violence, especially as the online environment removes the constraints of time and location on these behaviors \citep{lee_impact_2024}. On the media level, certain online media in their pursuit of higher click rates may even exacerbate the spread of cyber violence. At the user level, some netizens, lacking rational thought and moral restraint, are easily influenced by collective emotions, making them prone to participating in cyber violence. Due to the inherent ambiguity of the concept of cyber violence, the distinction between cybercrime and cyber violence remains unclear. This has led to a number of challenges in research related to the governance of cyber violence. It has been observed that cyber violence is a product of the Internet, and it can develop in parallel with the advancement of Internet technology, resulting in a definition that is dynamic and forms of expression that are continually evolving \citep{tao2025, luSentiment2024}. Based on previous research findings \citep{European,chan_cyberbullying_2021} and the extracted characteristics of cyber violence (see Appendix A.), this study defines the concept of cyber violence as follows:

Cyber violence refers to inappropriate behaviors conducted online through forms such as text, images, videos, and other mediums to attack, insult, abuse, or threaten others, causing direct or indirect harm to the victim. It manifests itself in a variety of forms, including but not limited to cyberbullying, online insults, online defamation, online rumors, hate speech, and cyber manhunt, etc \citep{song2021}. These behaviors often exhibit characteristics of anonymity, concealment, and rapid dissemination \citep{chugh_stalking_2022}, making it difficult for the victim to prevent and address.

\subsection{Research topics of cyber violence}
\rev{Governance of cyber violence} is increasingly drawing widespread attention from scholars, policymakers, and\rev{society}. Existing research mainly focuses on influencing factors, consequences, and intervention strategies.\rev{Cyber violence results from a complex interplay of multiple factors \citep{quynhho2022}, yet most studies examine either individual traits or social environments, lacking integration of micro and macro perspectives.} Individual factors include age, gender, personality, and daily Internet usage time \citep{mkhize_cyberbullying_2021},\rev{while macro factors involve limited social support \citep{arato_risk_2022} and weakened cultural norms. Thus, further exploration is needed to integrate micro and macro factors to improve the governance framework. Meanwhile, growing attention has been paid to the consequences of cyber violence. Longitudinal studies show that it can cause long-term psychological harm, such as depression, anxiety, and even suicidal tendencies \citep{camerini_cyberbullying_2020}.}

\rev{The above studies thoroughly explain the impacts and consequences of cyber violence on individuals. The cyber violence diffusion is a complex ecological process involving multi-agent strategic games and opinion evolution. For example, in virtual space, perpetrators reduce attack costs through anonymity. Their actions are transformed into public issues by online media's reporting, while public opinion pressure forces the government to adjust regulatory policies. Thus, focusing on multiple agents aids effective intervention design. The above studies preliminarily reveal game relations among agents. \citet{peter} revealed the game-theoretic relationship between victims and perpetrators, but overlooked the influence of external forces on the diffusion of cyber violence. \citet{glance2024cyberviolence} investigated the game relationship  between the government and perpetrators, and proposed intervention strategies such as legal sanctions and administrative regulation \citep{ng_effectiveness_2022, christner2025}; however, it failed to incorporate the interactions among diverse agents such as online media, making it difficult to reflect the complexity of the cyber violence ecosystem. \citet{xiao2019} included the triadic interaction among government, media, and netizens, but lacked a characterization of the behavior of victims, who are directly involved in the process.}\rev{From a theoretical view, \citet{mccombs2014new}’s agenda-setting theory shows media reports shape public perception, influencing victims’ willingness to defend rights (e.g., positive coverage increases perceived social support and lowers costs) and perpetrators’ risk expectations (e.g., exposure increases moral condemnation costs). Based on \citet{ostrom1994rules}’s public governance and social governance theories, government policy tools directly affect other actors’ game cost matrices. In view of the critical roles of the government and online media, and based on existing literature, this study constructs a four-party game model including the victim, perpetrator, online media, and government to better reflect reality. Meanwhile, the online media reporting directly affect the opinion ecology: objective reporting guides opinion positively \citep{helberger_political_2020}, while inaccurate or inflammatory reports \citep{budak_misunderstanding_2024} can lead to group polarization \citep{lian_cyber_2022}. Thus, the government, as a governance leader, needs to go beyond the single regulatory role and transform the online media into governance collaborators, thereby constructing a opinion governance system covering rule-making, behavioral adjustment, feedback, and optimization. Thus, this study focuses on the government and online media intervention strategies’ effects on opinion evolution during the dissemination stage.}

\rev{Moreover, while existing studies have proposed intervention strategies that alleviate cyber violence to some extent, they often overlook flexibility and specificity. For example, several studies advocated strict punishment \citep{piolanti_psychological_2022}, but failed to consider the dynamic match between punishment targets and severity,} which may even trigger a rebound effect \citep{Martinsen2022},\rev{reducing governance effectiveness. Although the necessity of diversified interventions has gained attention \citep{chu_game-theoretic_2024}, most current strategies rely on theoretical assumptions, lacking scenario-based simulations and case validation, raising concerns about their adaptability in complex settings. A coordinated governance mechanism is urgently needed to enhance flexibility and adaptability.}

\subsection{Research methods of cyber violence}

In the field of cyber violence research, existing survey \citep{zhou_association_2024} and experimental \citep{chan_cyberbullying_2021} methods have provided important empirical support for uncovering the prevalence, influencing factors, and consequences of cyber violence. However, these methods have certain limitations. Firstly, they often fail to fully account for the dynamic evolution of cyber violence within complex social network ecosystems. Secondly, existing analytical frameworks are largely confined to static correlation analysis or unidirectional causal inference, lacking the capacity to effectively capture the intricate interactions of cyber violence within multi-agent strategic dynamics and information diffusion processes. To address these limitations, scholars have turned to interdisciplinary approaches, such as
communication studies and computer science, to more comprehensively analyze the evolution of cyber violence and propose governance strategies. In the field of propagation dynamics modeling, \citet{kempe_maximizing_2003} proposed the Linear Threshold (LT) and Independent Cascade (IC) models to maximize the spread range by optimizing the selection of initial nodes. However, these models assume that node influence remains constant and that the propagation process is independent across interactions, making it difficult to characterize the dynamic interactive features of real social networks. Although \citet{yan_stochastic_2021} improved the IC model to simulate the diffusion of cyber violence, their assumption of a one-time opinion shift still simplifies opinion evolution into an irreversible process, resulting in static constraints on countermeasures. In contrast, the infectious disease SIR model allows individuals to transition dynamically between states, exhibiting the complex emergent properties of information dissemination. Essentially, it can be regarded as  generalization of the IC model \citep{chu_game-theoretic_2024}, providing a flexible framework for modeling the dynamic evolution of cyber violence.

Further research has revealed a strong coupling relationship between group polarization in online environments and emotion-driven mechanisms. Scholars have thus embedded emotional contagion mechanisms into infectious disease models, constructing emotion-enhanced frameworks for opinion propagation dynamics \citep{yuan2022, ma2024}. Regarding intervention strategies, \citet{chu_game-theoretic_2024} adopted a differential game model to explore unilateral governance strategies for counter-cyber violence information. However, such studies are limited to the decision-making of platform administrators and fail to analyze the collaborative mechanisms involving governments, social organizations, and other stakeholders. To address these limitations, this study innovatively incorporates the strategic interactions between governments and online media, introduces a bystander emotion parameter, and improves the SIR model to characterize the evolution of cyber violence-related public opinion. This choice is not only based on the extensively validated applicability of the SIR model in information dissemination research but also on its ability to flexibly simulate the coupling process of individual state transitions and strategic interactions. Furthermore, key factors such as bystanders, disseminators of pro-cyberviolence, disseminators of anti-cyberviolence, and intervention strategy combinations are comprehensively considered, making the model more consistent with real-world cyber violence scenarios.

In summary, this study integrates evolutionary game theory and communication dynamics to propose a two-stage, multi-scenario collaborative governance mechanism for cyber violence. Through simulation experiments and case studies, it analyzes the collaborative governance paths of government and online media during the game and public opinion dissemination stages, providing insights for formulating more adaptive governance strategies.

\section{Research design}

\begin{figure}[H]   
\centering
    \includegraphics[width=0.7\linewidth]{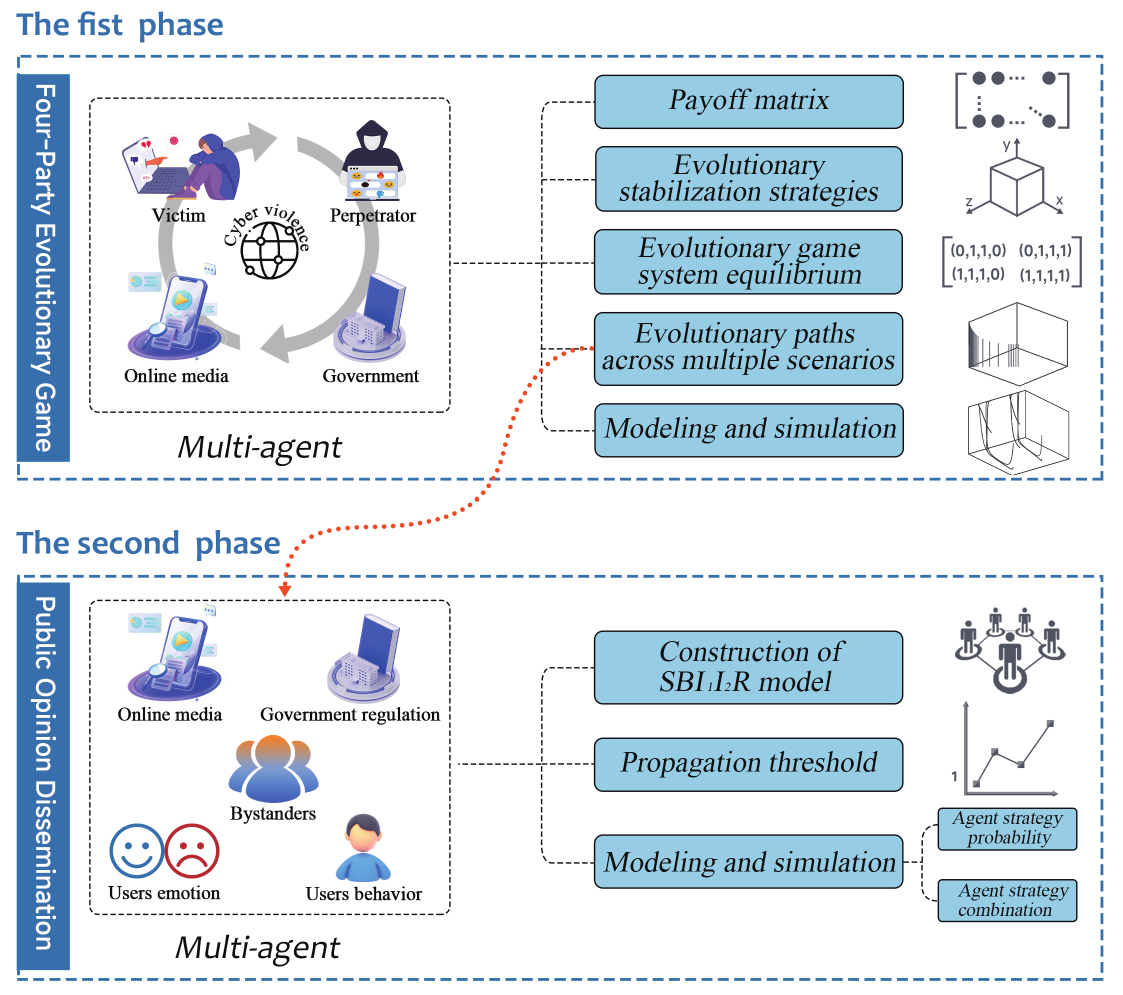}
    \caption{Framework of research design.}
    \label{fig:Framework of research design.}
\end{figure}

The governance of cyber violence is a long-term, comprehensive, and systematic endeavor that requires the joint participation of multiple stakeholders, including the government, the online media, and the public. Firstly, the governance of cyber violence involves a process where various societal stakeholders engage in dynamic strategic games to protect their interests. Secondly, as an open and collective public issue, cyber violence garners significant attention from Internet users, leading to the fermentation and dissemination of public opinion. The attitudes and behaviors of netizens are shaped by the strategies of online media and government actions, which shape the trajectory of public opinion. Based on this, the study divides the collaborative governance of cyber violence into two phases: the first is the evolutionary game phase of collaborative governance, and the second is the public opinion dissemination phase, built upon the foundation established in the first phase. Based on this, the study divides the collaborative governance of cyber violence into two phases: (1) the evolutionary game phase of collaborative governance; (2) the public opinion dissemination phase, which builds upon the foundation established in the first phase. The detailed research design is illustrated in \hyperref[fig:Framework of research design.]{Fig.~\ref*{fig:Framework of research design.}}.

This study collected 30 typical cases of cyber violence, drawn from 16 provinces, 3 municipalities, and 1 autonomous region in China, spanning the period from 2016 to 2024 (see Appendix A.). This study selects cases from China primarily because the country is intensifying its efforts to combat cyber violence  \citep{ChinaLawNews}. Several Chinese public security agencies have disclosed related cases and penalties, providing a foundation for the study. Additionally, China’s specific legislation on cyber violence is still in the exploratory stage, this study offers insights for other countries facing similar challenges and in urgent need of scientific decision-making guidance. The selected cases meet two criteria: (1) sparking widespread social discussion and generating relevant public opinion; (2) being officially disclosed by Chinese government departments. For example, typical cases of cyber violence violations and crimes disclosed by the Supreme People’s Court of the People’s Republic of China in 2023 \citep{ThePressOffice}, cyber violence violations and crimes disclosed by the Ministry of Public Security of China in 2023 \citep{MinistryofPublic}, and typical cases of cyber violence violations and crimes disclosed by provincial or municipal public security agencies. Through an in-depth analysis of these cases, major players in cyber violence incidents, namely the victim, perpetrator, the online media, and the government, are identified. These players serve as the foundation for evolutionary game phase. Furthermore, these cases have sparked widespread public opinion and attracted international media attention. Notably, cyber violence, as a global issue, possesses universal characteristics in its nature, manifestations, and impacts. Although based on Chinese cases, this study’s theoretical framework and modeling methods, with appropriate adjustments, have cross-cultural applicability and value.

\subsection{Description of the problem}
\subsubsection{Four-party evolutionary game phase}

This study introduces an evolutionary game model in the multi-agent game phase to examine the behavioral strategies and interactive impacts among the victim, the perpetrator, the online media, and the government. The definitions of the four stakeholders in the collaborative governance of cyber violence are as follows:

(1) The victim

The victim is the direct recipient of cyber violence, who may become targets due to their statements, actions, or identity.

(2) The perpetrator

The perpetrator is the initiator of cyber violence, who may publish offensive statements or engage in aggressive behavior driven by various motivations, such as personal grudges, conflicts of interest, or entertainment purposes, to achieve their objectives.

(3) The online media

Online media refers to news outlets and self-media on platforms like Sina Weibo, WeChat, Douyin (the Chinese version of TikTok), forums, or websites, excluding official affiliates. Their stances significantly influence the spread of cyber violence public opinion. Correct guidance involves conveying accurate values through in-depth investigations, objective reporting, and rational commentary, fostering a transparent and healthy online atmosphere. Misguidance, on the other hand, spreads inaccurate or misleading information \citep{xie_information_2020}, leading to misunderstandings, biases, and inappropriate behaviors, which can exacerbate cyber violence and hinder a healthy online environment.

(4) The government 

The term “government” refers to state administrative institutions that establish and officially verify social media accounts on various online platforms to manage public affairs. As a key leader and participant in cyberspace, its attitude and strategic actions critically influence the dissemination of public opinion on cyber violence. It uses official accounts to disseminate policy information \citep{mansoor_citizens_2021} convey official positions, and guide public opinion. For instance, the government is responsible for regulating online activities \citep{helberger_political_2020}, penalizing illegal actions, and issuing public reports. It also engages in extensive interactions with other stakeholders to enhance public service quality and collectively maintain a healthy and civilized cyberspace.

Government departments, with varying governance philosophies and approaches, typically adopt two regulatory strategies: “strong regulation” and “weak regulation,” which are defined below.

Strong regulation by the government refers to an active, proactive, and stringent approach adopted by government departments in the governance of cyber violence. This approach includes but is not limited to, the following measures: (a) frequent updates on government affairs; (b) strict guidance of public opinion; (c) severe crackdown on illegal behaviors; (d) active interaction and feedback with stakeholders. 
Weak regulation by the government refers to a relatively lenient and passive approach adopted by government departments in the governance of cyber violence. This approach includes but is not limited to, the following measures: (a) infrequent updates on government affairs; (b) lack of public opinion guidance; (c) insufficient penalties; (d) limited interaction and feedback with stakeholders. 

Notably, “strong regulation” and “weak regulation” are not absolute concepts but relative ones. In practice, the government departments should flexibly adjust their regulatory strategies and severity based on the specific circumstances of cyber violence and societal needs, to foster the healthy and orderly development of cyberspace.
\subsubsection{Public opinion dissemination phase}
\rev{One of the core challenges in the current governance of cyber violence lies in the insufficient collaborative effectiveness of public intervention mechanisms. Therefore, this phase focuses on the influence mechanisms of intervention strategies adopted by the online media and the government on the evolution of public opinion in cyber violence.} In the context of cyber violence, bystanders are individuals who witness acts of cyber violence but do not directly participate or explicitly express support or opposition. The attitudes and behaviors of bystanders influence the occurrence, spread, and evolution of public opinion of cyber violence. Accordingly, this study incorporates bystanders into the traditional infectious disease model to construct an optimized $SB{{I}_{1}}{{I}_{2}}R$  public opinion dissemination model. \rev{The model specifically examines the effects of intervention strategies by the online media and the government on the states of netizens and incorporates emotional interactions among users, thereby providing a more realistic representation of the dynamic evolution patterns of cyber violence public opinion.}

\rev{It should be noted that the public opinion dissemination  model does not yet include the victim and the perpetrator as strategic analysis units. This does not negate the importance of individuals, but rather focuses on the key mechanisms of public governance agents. The behaviors of the victim and the perpetrator at the public opinion dissemination phase can still be understood from the evolutionary game phase.  For example, when the related behaviors of the victim and the perpetrator trigger a surge of public attention, the momentum of public opinion creates an external pressure field that influences the strategic choices of the government and online media. Meanwhile, the behavioral strategies of online media and the government will further affect the decisions of the victim and the perpetrator. For instance, when the probability of “strong regulation” by the government or “correct guidance” by online media increases, the victim tends to adopt a “non-action” strategy once their rights are sufficiently protected, while the perpetrator tends to adopt a “stop attacking” strategy as the cost of violations rises (as indicated by \hyperref[3]{Eq.(3)} and \hyperref[7]{Eq.(7)})}

\subsection{Model assumptions }
\subsubsection{Basic assumptions of the four-party evolutionary game model}
\label{sec:four-party evolutionary game model}

\textbf{Hypothesis 1.} Given the information asymmetry and varying perceived benefits a cyber violence outbreak, stakeholders are boundedly rational and unable to achieve optimal strategies at the game’s start. They adjust their strategies based on observations of others, eventually reaching a stable equilibrium state known as an evolutionary stable strategy.

\textbf{Hypothesis 2.} In the complex cyberspace, stakeholders, constrained by bounded rationality and incomplete information, will adopt varying behavioral strategies. The strategy space for the victim is \{action, non-action\}. The $x$ denotes the probability that the victim adopts the “action” strategy, while $1-x$ denotes the probability of choosing “ non-action,” with $x\in [0,1]$. The strategy space for the perpetrator is \{keep attacking, stop attacking\}. The $y$ denotes the probability that the perpetrator adopts the “stop attacking” strategy, while $1-y$ denotes the probability of choosing “keep attacking,” with $y\in [0,1]$. The strategy space for the online media is \{correct guidance, misguidance\}. The $z$ denotes the probability that the online media adopts the “correct guidance” strategy, while $1-z$ denotes the probability of opting for “misguidance,” with $z\in[0,1]$. The strategy space for the government is \{strong regulation, weak regulation\}. The $m$ denotes the probability that the government adopts the “strong regulation” strategy, while $1-m$ denotes the probability of choosing “weak regulation,” with $m\in [0,1]$.

\textbf{Hypothesis 3.} When the victim adopts the “action” strategy, they need to bear the costs (denoted as ${{C}_{v1}}$) associated with exposing the situation, reporting to the police, and pursuing legal proceedings. Simultaneously, the victim may also gain benefits (denoted as ${{R}_{v1}}$), including safeguarding their rights, obtaining social support, and prompting legal sanctions against the perpetrator. During the process of experiencing cyber violence, the victim may endure the reopening of emotional wounds, breaches of privacy, and significant expenditures of time and energy. Many victims may lack the courage to expose the perpetrator’s malicious actions. When the victim adopts the “non-action” strategy, they must endure certain costs (denoted as ${{C}_{v2}}$) of cyber violence, including psychological pressure and negative emotions. In this scenario, if the online media adopts a “misguidance” strategy by publishing or reposting unverified and biased opinions, it may inflict additional losses (denoted as ${{H}_{1}}$) on the victim, including exposure to privacy, increased psychological stress, etc.; when the online media opts for “correct guidance,” it can gain certain benefits (denoted as $W$) of career fulfillment; if the government adopts a “weak regulation” strategy, the victim may suffer additional losses (denoted as ${{H}_{2}}$),\rev{including but not limited to increased costs of evidence collection and diminished availability of social support networks;} if the perpetrator adopts “keep attacking” strategy, \rev{the victim will incur additional losses (denoted as $ {{H}_{3}}$), including but not limited to aggravated psychological trauma, prolonged recovery duration and diminished social trust.}

\textbf{Hypothesis 4.} The perpetrator seeks to fulfill their psychological needs or objectives by publishing or reposting extreme and negative views \citep{lian_cyber_2022}. When the perpetrator opts for the “stop attacking” strategy, they may gain potential benefits (denoted as ${{R}_{p}}$), such as personal growth, psychological recovery, mended relationships, and the rebuilding of trust and recognition in society. When the perpetrator adopts to “keep attacking” strategy, they incur the costs (denoted as ${{C}_{p}}$) of creating and disseminating information about cyber violence. In this scenario, if the online media opts for “misguidance,” it may provide potential benefits (denoted as $\delta$) to the perpetrator, such as psychological satisfaction; when the victim takes “action,” it can result in losses (denoted as ${{L}_{p}}$) for the perpetrator, such as the loss of potential satisfaction, influence, and increased risk of punishment; when the government implements “strong regulation,” the perpetrator will face certain penalties (denoted as ${{g}_{1}}$), including legal sanctions and fines. The severity of the punishment is denoted as $\phi$$(0\le \phi \le 1)$ . 

\textbf{Hypothesis 5.} When a cyber violence incident escalates and enters the public domain, online media may adopt various strategies within the strategy space \{correct guidance, misguidance\}. Due to professional ethics and government oversight, some online media will choose the correct guidance strategies. However, certain outlets, seeking to generate hype and drive traffic, may exploit regulatory loopholes and opt for “misguidance” strategies, reflecting a risk-taking mindset \citep{ farrar_punish_2023}. The operational cost incurred by the online media that choose “misguidance” is denoted as ${{C}_{m2}}$. Although online media may achieve higher speculative benefits (denoted as ${{R}_{m2}}$), such as increased traffic and attention in the short term. Nevertheless, in the long term, this will harm their reputation, with the reputation loss denoted as $P$. Additionally, they may face penalties (denoted as ${{g}_{2}}$) from government authorities, such as rectification orders and fines, with the severity of the punishment denoted as $\beta $$(0\le \beta \le 1)$. When the online media choose the “correct guidance” strategy, they incur an operational cost denoted as ${{C}_{m1}}$(${{C}_{m2}}<{{C}_{m1}})$. It includes expenses such as the manpower and resources required for investigation and verification when condemning the perpetrator. If the perpetrator chooses to “stop attacking” at this point, the online media will benefit (denoted as ${{R}_{m1}}$) from increased traffic, enhanced public favorability, greater visibility, etc.

\textbf{Hypothesis 6.} In response to a cyber violence incident, the government’s strategies will be continuously revised and adjusted based on considerations of public trust, image, political performance, and expected benefits. When the government adopts the “strong regulation” strategy, it incurs costs (denoted as ${{C}_{g1}}$) in terms of manpower, resources, and time \citep{European}. At the same time, if the victim takes the “action” strategy, it will be accompanied by social benefits (denoted as ${{F}_{1}}$),\rev{such as reducing the population of potential victims through deterrence mechanisms, elevating collective consciousness of rights protection, and promoting social justice;} if the online media adopts the “correct guidance” strategy, it will yield benefits (denoted as ${{F}_{2}}$), such as social stability and enhanced public credibility. When the government adopts the “weak regulation” strategy, the associated cost is denoted as ${{C}_{g2}}$(${{C}_{g2}}<{{C}_{g1}})$. If the perpetrator opts to “keep attacking,” it will incur corresponding governance costs or losses (denoted as ${{F}_{3}})$, including social disorder and diminished public credibility.

\rev{Among these, the punishment severity $\phi$ and $\beta $ are set within the range of [0,1], where 0 represents no punitive measures and 1 denotes the maximum level of punishment, consistent with the approaches adopted in \citet{liEvolutionaryGame2015}'s research on evolutionary game models of rumor spreading and \citet{geng_online_2023}'s study on public opinion governance.} A table of specific parameters and their meanings is provided in Appendix B. for easy reference.

%% Use a table environment to create tables.
%% Refer following link for more details.
%% https://en.wikibooks.org/wiki/LaTeX/Tables

\subsubsection{Basic assumptions of the public opinion dissemination model}
\label{sec:Basic assumptions of the public opinion}
\textbf{Hypothesis 1.}  $S$ (Susceptible): The susceptible individuals who have not yet been exposed to public opinion information on Internet platforms; $B$ (Bystanders): Bystanders who are exposed to public opinion information and remain in an observational state without expressing their opinions; $I_1$ (Infected 1): The disseminators who are against cyber violence (abbreviated as “disseminators of anti-cyberviolence”, hereinafter referred to in the same manner) are individuals infected by the attitude of cyber violence and hold opposing attitudes or views on cyber violence. For example, individuals who post or repost positive comments about anti-cyberviolence; $I_2$ (Infected 2): The disseminators who support cyber violence (abbreviated as “disseminators of pro-cyberviolence”, hereinafter referred to in the same manner) are individuals infected by the attitude of cyber violence and hold supporting attitudes or views on cyber violence. For example, individuals who post or repost negative comments about pro-cyberviolence; $R$ (Recovered): Immunized individuals who are not influenced by the attitudes of cyber violence.

\textbf{Hypothesis 2.} Typically, during the initial stage of public opinion development, online media guides netizens by presenting diverse viewpoints. Government intervention occurs only when the number of supporters of cyber violence reaches a certain threshold. Given this, it is assumed that the online media intervention primarily affects the transition from state $B$ to state $I$, while the government, by virtue of its credibility, can influence both the transition from state $I$ to state $B$ and the transition from state $I$ to state $R$.

\textbf{Hypothesis 3.}\rev{Internet platforms serve as open environments for public discourse, with user populations undergoing continual renewal and expansion. In this study, the input rate $A$ is introduced to denote the rate of new users joining the system. High-speed networks and distributed computing have greatly lowered access barriers and markedly enhanced real-time responsiveness and accessibility, enabling large-scale user influxes over short time spans. Particularly during cyber violence incidents, the immediacy of information dissemination and the emotional contagion effect often drive user participation rates to approach the system’s upper limit. To simulate such real-world scenarios, this study assumes an initial input rate $A$ of 1.} The probability of transitioning from state $S$ to state $B$ is denoted as $\sigma$. The immunity probability of netizens transitioning from  state $S$ to state $R$ due to a lack of interest in or disbelief in public opinion information is denoted as $\theta$. In the initial stage, the online media intervenes by disseminating either correct or incorrect viewpoints. Bystanders influenced by correct or incorrect information have a certain probability of becoming infected with either type $I_1$ or type $I_2$, where ${{I}_{1}}+{{I}_{2}}=1$. Netizens are influenced by both the correct guidance of the online media and emotional identification. The probability of transitioning from state $B$ to $I_1$ is $({{z}_{1}}+{{a}_{1}})$, where ${z}_{1}$ represents the probability of correct guidance from the online media, and ${a}_{1}$ is the parameter for positive emotional identification. When the online media adopts “misguidance,” the probability of netizens transitioning from state $B$ to $I_2$ is $({{z}_{2}}+{{a}_{2}})$, where ${z}_{2}$ represents the probability of the online media providing “misguidance,” and ${a}_{2}$ is the parameter for negative emotional identification.\rev{Where $z_{1}$ + $z_{2}$ =1.}

\textbf{Hypothesis 4.} Only when the government implements “strong regulation” will the netizens transition from state $I$ to $R$. When the government adopts “strong regulation,” netizens are influenced by both the government’s “strong regulation” and emotional identification. The probability of netizens transitioning from state $I_1$ to $R$ is $k({{m}_{1}}+{{n}_{1}})$ , and the probability of transitioning from state $I_2$ to $R$ is $k({{m}_{1}}+{{n}_{2}})$. Where $k$ represents the government’s credibility \citep{park_policy_2016},  ${m}_{1}$ is the probability of the government’s “strong regulation.” ${n}_{1}$ and ${n}_{2}$ are both positive emotional identification parameters $({{n}_{2}}<{{n}_{1}})$. When the government adopts “weak regulation,” it cannot effectively prevent netizens from expressing negative opinions, i.e., the probability of transitioning from state $I_2$ to $B$ is $k{{m}_{2}}$, where ${m}_{2}$ is the probability of the government’s weak regulation.\rev{Where $m_{1}$ + $m_{2}$ =1.}

\subsection{Model construction and solution}
\subsubsection{Construction and solution analysis of evolutionary game models}
\begin{figure}[H]
    \centering
    \includegraphics[width=0.65\linewidth]{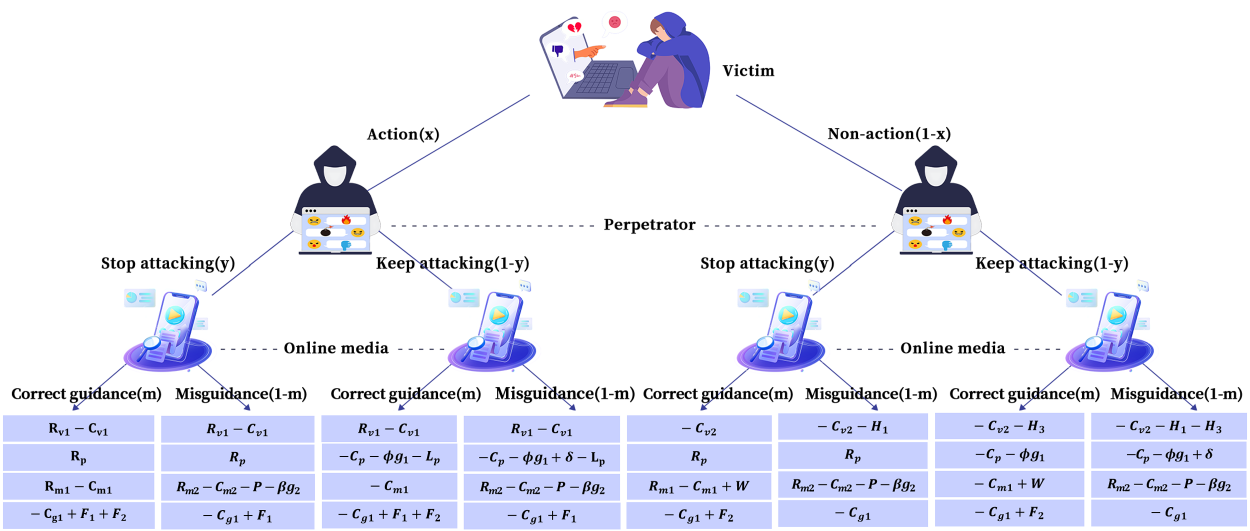}
     \caption{{Payoff Matrix when the government adopts the “strong regulation” strategy.}}
    \label{fig:Payoff Matrix when the government adopts the “strong regulation” strategy}
\end{figure}

\begin{figure}[H]
    \centering
    \includegraphics[width=0.66\linewidth]{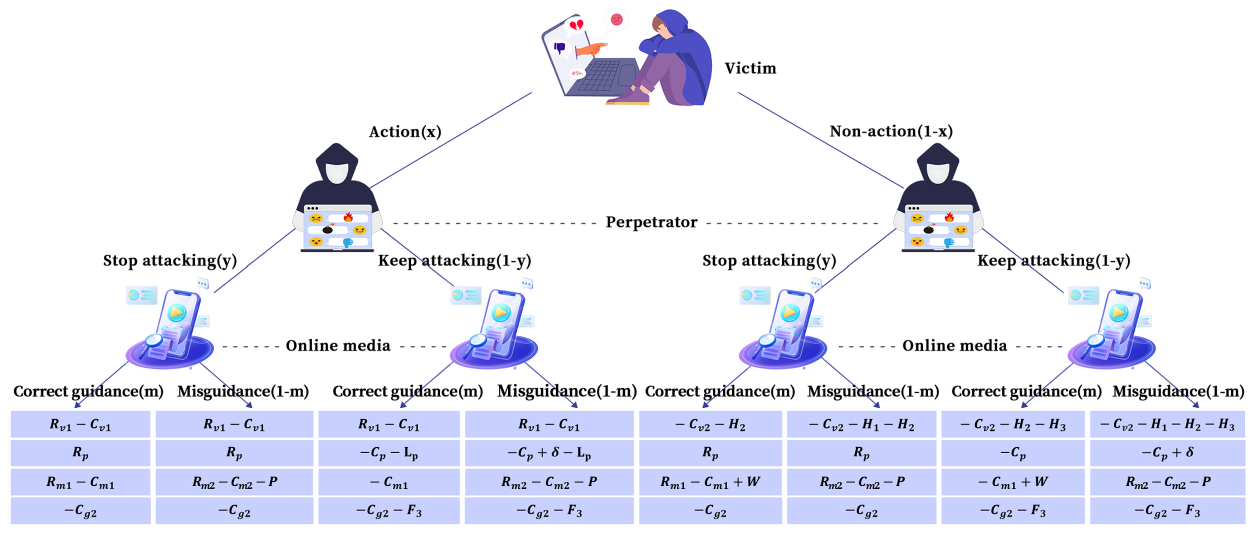}
     \caption{{Payoff Matrix when the government adopts the “weak regulation” strategy.}}
    \label{fig:Payoff Matrix when the government adopts the “weak regulation” strategy}
\end{figure}

Based on the above assumptions, the payoff matrix for the evolutionary game model of collaborative governance of cyber violence under different government behavior strategies can be calculated (see in \hyperref[fig:Payoff Matrix when the government adopts the “strong regulation” strategy]{Fig.~\ref*{fig:Payoff Matrix when the government adopts the “strong regulation” strategy}} and \hyperref[fig:Payoff Matrix when the government adopts the “weak regulation” strategy]{Fig.~\ref*{fig:Payoff Matrix when the government adopts the “weak regulation” strategy}}).

(1) Replication dynamic equations and evolutionary
stabilization strategies for stakeholders

(a) Replication dynamic equations and evolutionary stabilization strategies for the victim

The expected payoffs for the victim adopting the “action” or “non-action” strategy, the replicator dynamic equation for the victim’s strategy, and the first derivative are as follows:
\vspace{-0.5em}
\begin{equation}
\vspace{-1em}
U_x = Rv_{1} - Cv_{1}
\end{equation}
\begin{equation}
{{U}_{(1-x)}}={{H}_{2}}m-{{C}_{v2}}-{{H}_{1}}-{{H}_{2}}-{{H}_{3}}+{{H}_{3}}y+{{H}_{1}}z
\end{equation}
\begin{equation}
F(x)=dx/dt=x({{U}_{x}}-\overline{U})=x(1-x)({{R}_{v1}}-{{C}_{v1}}-{{H}_{2}}m+{{C}_{v2}}+{{H}_{1}}+{{H}_{2}}+{{H}_{3}}-{{H}_{3}}y-{{H}_{1}}z)
\label{3}
\end{equation}
\begin{equation}
{F}'(x)=(1-2x)({{R}_{v1}}-{{C}_{v1}}-{{H}_{2}}m+{{C}_{v2}}+{{H}_{1}}+{{H}_{2}}+{{H}_{3}}-{{H}_{3}}y-{{H}_{1}}z)
\end{equation}

Under the stability theorems and properties of differential equations, when $F(x)=0$, ${F}'(x)<0$, the stable point can be obtained.

\textbf{Proposition 1.}  When $m>{{m}_{0}}$, the stable strategy for the victim is “non-action”. When $m<{{m}_{0}}$, the stable strategy is “action.” When $m={{m}_{0}}$, the stable strategy cannot be determined. Where the threshold value is
${{m}_{0}}=({{R}_{v1}}-{{C}_{v1}}+{{C}_{v2}}+{{H}_{1}}+{{H}_{2}}+{{H}_{3}}-{{H}_{3}}y-{{H}_{1}}z)/{{H}_{2}}$.

(b) Replication dynamic equations and evolutionary stabilization strategies for the perpetrator

The expected payoffs for the perpetrator adopting the “keep attacking” or “stop attacking” strategy, the replicator dynamic equation for the perpetrator’s strategy, and the first derivative are as follows:	
\vspace{-0.5em}
\begin{equation}
{{U}_{y}}={{R}_{p}}
\vspace{-1em}
\end{equation}
\begin{equation}
{{U}_{1-y}}=-\delta z+\delta -{{C}_{p}}-\phi {{g}_{1}}m-{{L}_{p}}x
\end{equation}
\begin{equation}
F(y)=dy/dt=y({{U}_{y}}-\overline{U})=y(1-y)({{R}_{p}}+\delta z-\delta +{{C}_{p}}+\phi {{g}_{1}}m+{{L}_{p}}x)
\label{7}
\end{equation}
\begin{equation}
{F}'(y)=(1-2y)({{R}_{p}}+\delta z-\delta +{{C}_{p}}+\phi {{g}_{1}}m+{{L}_{p}}x)
\end{equation}

The same as above, when $F(y)=0$, ${F}'(y)<0$, the stable point can be obtained.

\textbf{Proposition 2.} When $m>{{m}_{0}}$, the stable strategy for the perpetrator is “stop attacking”. When $m<{{m}_{0}}$, the stable strategy is “keep attacking”. When $m={{m}_{0}}$, the stable strategy cannot be determined. Where the threshold value is
${{m}_{0}}=({{R}_{p}}+\delta z-\delta +{{C}_{p}}+{{L}_{p}}x)/-\phi {{g}_{1}}$.

(c) Replication dynamic equations and evolutionary stabilization strategies for the online media

The expected payoffs for the online media adopting the “correct guidance” or “misguidance” strategy, the replicator dynamic equation for the online media strategy, and the first derivative are as follows:
\vspace{-0.5em}
\begin{equation}
{{U}_{z}}=W-{{C}_{m1}}+{{R}_{m1}}y-Wx
\vspace{-1em}
\end{equation}
\begin{equation}
{{U}_{1-z}}=-P-{{C}_{m2}}+{{R}_{m2}}-\beta {{g}_{2}}m
\end{equation}
\begin{equation}
F(z)=dz/dt=z({{U}_{z}}-\overline{U})=z(1-z)(W-{{C}_{m1}}+{{R}_{m1}}y-Wx+P+{{C}_{m2}}-{{R}_{m2}}+\beta {{g}_{2}}m)
\end{equation}
\begin{equation}
{F}'(z)=(1-2z)(W-{{C}_{m1}}+{{R}_{m1}}y-Wx+P+{{C}_{m2}}-{{R}_{m2}}+\beta {{g}_{2}}m)\
\end{equation}

The same as above, when $F(z)=0$, ${F}'(z)<0$, the stable point can be obtained.

\textbf{Proposition 3.}  When  $m>{{m}_{0}}$, the stable strategy for the online media is “correct guidance”. When  $m<{{m}_{0}}$, the stable strategy is “misguidance”. When  $m={{m}_{0}}$, the stable strategy cannot be determined. Where the threshold value is  ${{m}_{0}}=(W-{{C}_{m1}}+{{R}_{m1}}y-Wx+P+{{C}_{m2}}-{{R}_{m2}})/-\beta {{g}_{2}}$.

(d) Replication dynamic equations and evolutionary stabilization strategies for the government

The expected payoffs for the government adopting the “strong regulation” or “weak regulation” strategy, the replicator dynamic equation for the government’s strategy, and the first derivative are as follows:
\vspace{-0.5em}
\begin{equation}
{{U}_{m}}=-{{C}_{g1}}+{{F}_{1}}x+{{F}_{2}}z
\vspace{-1em}
\end{equation}
\begin{equation}
{{U}_{1-m}}=-{{C}_{g2}}+{{F}_{3}}y-{{F}_{3}}
\end{equation}
\begin{equation}
F(m)=dm/dt=m({{U}_{m}}-\overline{U})=m(1-m)(-{{C}_{g1}}+{{F}_{1}}x+{{F}_{2}}z+{{C}_{g2}}-{{F}_{3}}y+{{F}_{3}})
\end{equation}
\begin{equation}
{F}'(m)=(1-2m)(-{{C}_{g1}}+{{F}_{1}}x+{{F}_{2}}z+{{C}_{g2}}-{{F}_{3}}y+{{F}_{3}})
\end{equation}

The same as above, when $F(m)=0$, ${F}'(m)<0$, the stable point can be obtained.  

\textbf{Proposition 4.} When  $y>{{y}_{0}}$, the stable strategy for the government is “ weak regulation”. When $y<{{y}_{0}}$, the stable strategy is “ strong regulation”. When $y={{y}_{0}}$, the stable strategy cannot be determined. Where the threshold value is  ${{y}_{0}}=(-{{C}_{g1}}+{{F}_{1}}x+{{F}_{2}}z+{{C}_{g2}}+{{F}_{3}})/{{F}_{3}}$.

Proofs and explanations for Propositions 1 to 4 are provided in Appendix C.

(2) Evolutionary game system equilibrium analysis

Based on the replicator dynamic equations of the four-party game participants, the Jacobian matrix of the replicator dynamic system is obtained as follows:

\begin{equation}
J=\left[ \begin{matrix}
   \frac{\partial F(x)}{\partial x} & \frac{\partial F(x)}{\partial y} & \frac{\partial F(x)}{\partial z} & \frac{\partial F(x)}{\partial m}  \\
   \frac{\partial F(y)}{\partial x} & \frac{\partial F(y)}{\partial y} & \frac{\partial F(y)}{\partial z} & \frac{\partial F(y)}{\partial m}  \\
   \frac{\partial F(z)}{\partial x} & \frac{\partial F(z)}{\partial y} & \frac{\partial F(z)}{\partial z} & \frac{\partial F(z)}{\partial m}  \\
   \frac{\partial F(m)}{\partial x} & \frac{\partial F(m)}{\partial y} & \frac{\partial F(m)}{\partial z} & \frac{\partial F(m)}{\partial m}  \\
\end{matrix} \right]
\end{equation}

By setting $F(x)=0$, $F(y)=0$, $F(z)=0$ and $F(m)=0$, the local equilibrium points of the system can be calculated, yielding 16 pure strategy equilibrium points: A(0,0,0,0), B(1,0,0,0), C(0,1,0,0), D(0,0,0,1), E(1,1,0,0), F(1,0,0,1), L(0,1,0,1), G(1,1,0,1), M(0,0,1,0), N(1,0,1,0), H(0,1,1,0), O(0,0,1,1), P(1,0,1,1), I(0,1,1,1), J(1,1,1,0) and K(1,1,1,1). The eigenvalues of the 16 pure strategy equilibrium points are shown in Appendix D.

The stable solution in evolutionary games is a strict Nash equilibrium, and a strict Nash equilibrium must be a pure strategy \citep{friedman_evolutionary_1991}. Therefore, this study will analyze the stability of the 16 pure strategy equilibrium points in the four-party evolutionary game. Through the analysis of the above replicator dynamic equations, the stability of strategy combinations in the four-dimensional dynamical system of the evolutionary game for collaborative governance of cyber violence can be determined according to Lyapunov’s first method. When all the eigenvalues of the Jacobian matrix are less than 0, the equilibrium point is an ESS point. 11 equilibrium points may become ESS points, namely A, B, C, D, E, F, G, H, I, J, and K (see Appendix D.).

In the collaborative governance of cyber violence, online media and the government each play a crucial role, with four possible strategy combinations: \{misguidance, weak regulation\}, \{misguidance, strong regulation\}, \{correct guidance, weak regulation\}, and \{correct guidance, strong regulation\}.  Assuming the least ideal scenario, online media and the government would initially adopt the “misguidance” and “weak regulation,” strategies due to operational cost considerations.  As the situation evolves, their strategies gradually adjust, ultimately converging toward the optimal combination: \{correct guidance, strong regulation\}. To better understand this evolution process, we can further extend these four strategy combinations into four specific scenarios as follows:

\textbf{Scenario 1.} When online media adopts “misguidance,” the government initially implements a “weak regulation” strategy and shifts to a “strong regulation” strategy in the later stages.

\textbf{Scenario 2.} When online media adopts “correct guidance,” the government initially implements a “weak regulation” strategy and shifts to a “strong regulation” strategy in the later stages.

\textbf{Scenario 3.} When the government implements “weak regulation,” the online media initially adopts a “misguidance” strategy and shifts to a “correct guidance” strategy in the later stages.

\textbf{Scenario 4.} When the government implements “strong regulation,” the online media initially adopts a “misguidance” strategy and shifts to a “correct guidance” strategy in the later stages.

\begin{figure}[H]
    \centering
    \includegraphics[width=0.8\linewidth]{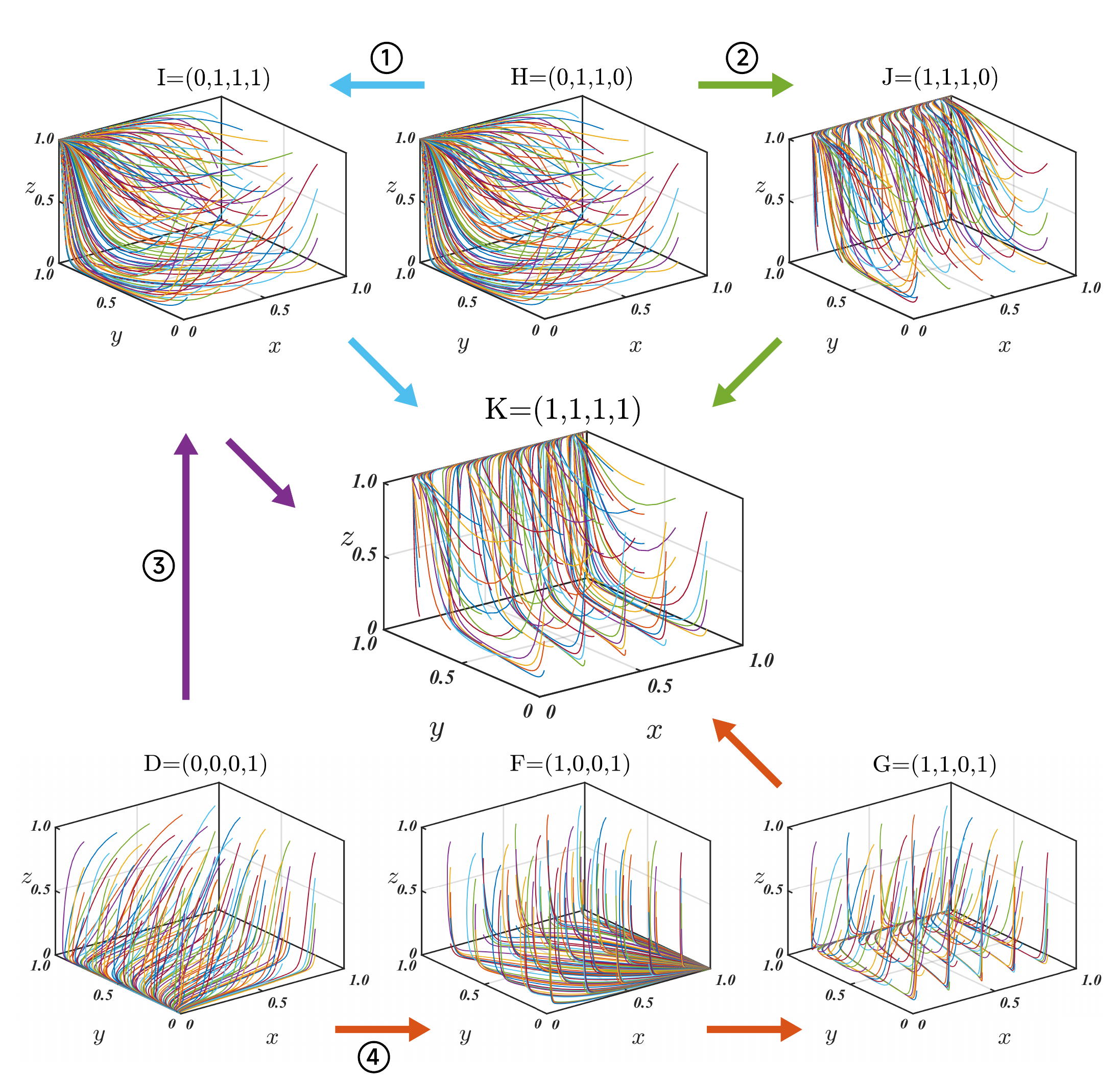}
    \caption{Evolution and path of 7 equilibrium points under different initial probabilities.}
    \label{fig:Evolution and path of 7 equilibrium points under different initial probabilities.}
\end{figure}

Due to space constraints, specific scenario analysis can be found in Appendix E. We can learn from it that, to reach the optimal equilibrium point K, there are four feasible paths in Scenarios 2 and 4 : H-J-K, H-I-K, D-I-K and D-F-G-K. In contrast, there are no ideal paths in Scenarios 1 and 3.  Then we conduct simulation experiments on the seven key equilibrium points involved in these four paths: H, J, D, F, G, I, and K. The replicator dynamic equations are allowed to evolve over 50 iterations, with the simulation results shown in \hyperref[fig:Evolution and path of 7 equilibrium points under different initial probabilities.]{Fig.~\ref*{fig:Evolution and path of 7 equilibrium points under different initial probabilities.}}. In this simulation, the parameter values for equilibrium point K (1,1,1,1) are used as the initial values, as detailed in \hyperref[sec:Parameter]{Section 4.1.1}. Parameter settings on simulation experiments. The values for other equilibrium points are adjusted from these initial values to satisfy the corresponding mathematical logic conditions.

As shown in \hyperref[fig:Evolution and path of 7 equilibrium points under different initial probabilities.]{Fig.~\ref*{fig:Evolution and path of 7 equilibrium points under different initial probabilities.}}, regardless of the initial strategy probabilities of the game subjects, the system ultimately converges to the corresponding seven equilibrium points. The simulation results are consistent with the conclusions drawn from the system’s evolutionary stability analysis, thereby validating the effectiveness of the model. Addtionally, we analyze the necessary real-world conditions required for the feasible paths (See Appendix F.).

\subsubsection{Construction and solution analysis of $SB{{I}_{1}}{{I}_{2}}R$ public opinion dissemination model}
(1) Model construction

\begin{figure}[http]
    \centering
    \includegraphics[width=0.6\linewidth]{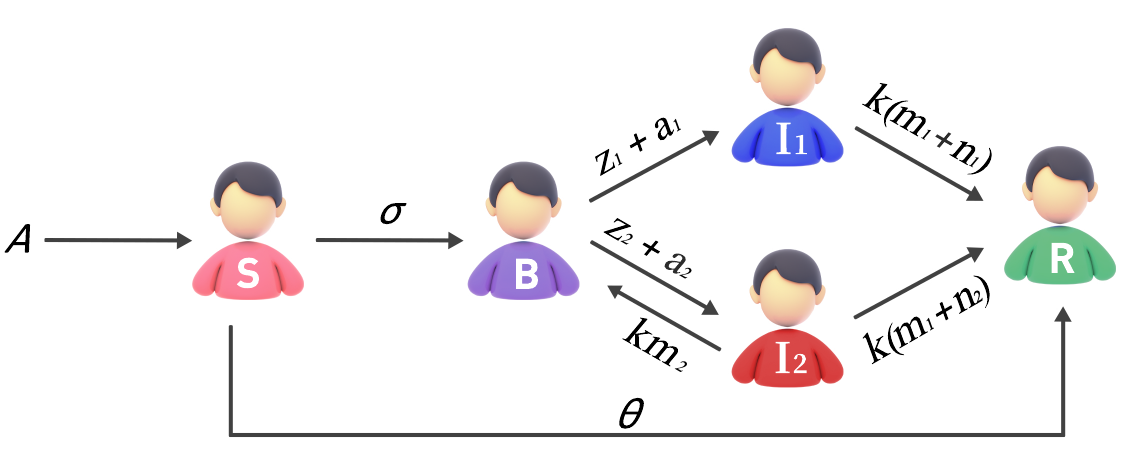}
     \caption{{Transformation rules of $SB{{I}_{1}}{{I}_{2}}R$ public opinion dissemination model.}}
    \label{fig:Transformation-rules}
\end{figure}

The conversion rules of the $SB{{I}_{1}}{{I}_{2}}R$ public opinion dissemination model can be derived from the above, as shown in \hyperref[fig:Transformation-rules]{Fig.~\ref*{fig:Transformation-rules}}.

Based on the theory of propagation dynamics, ordinary differential equations are constructed. According to \hyperref[fig:Transformation-rules]{Fig.~\ref*{fig:Transformation-rules}}, the $SB{{I}_{1}}{{I}_{2}}R$ epidemic dynamics equations for cyber violence public opinion, referred to as the System $Q$, can be expressed as follows:
\vspace{-0.5em}
\begin{equation}
\left\{ \begin{array}{*{35}{l}}
   {dS}/{dt}=A-\sigma S{{I}_{1}}-\sigma S{{I}_{2}}-\theta S  \\
   {dB}/{dt}=\sigma S{{I}_{1}}+\sigma S{{I}_{2}}-({{z}_{1}}+{{a}_{1}})B-({{z}_{2}}+{{a}_{2}})B+k{{m}_{2}}{{I}_{2}}  \\
   {d{{I}_{1}}}/{dt}=({{z}_{1}}+{{a}_{1}})B-k({{m}_{1}}+{{n}_{1}}){{I}_{1}}  \\
   {d{{I}_{2}}}/{dt}=({{z}_{2}}+{{a}_{2}})B-k({{m}_{1}}+{{n}_{2}}){{I}_{2}}-k{{m}_{2}}{{I}_{2}}  \\
   {dR}/{dt}=\theta S+k({{m}_{1}}+{{n}_{1}}){{I}_{1}}+k({{m}_{1}}+{{n}_{2}}){{I}_{2}}  \\
   N(t)=S(t)+B(t)+{{I}_{1}}(t)+{{I}_{2}}(t)+R(t)  \\
\end{array} \right.
\end{equation}

(2) Propagation threshold analysis

Using the next-generation matrix method, the propagation threshold ${{R}_{0}}$ (basic reproduction number) of the system $Q$ can be calculated. Let $X={{(S,B,{{I}_{1}},{{I}_{2}},R)}^{T}}$, then:
\vspace{-0.5em} % 调整此值来控制公式与前后文本的间距
\begin{equation}
{dX}/{dt}=F(X)-V(X)
\vspace{-0.5em}
\end{equation}

 ${{R}_{0}}=\rho (F{{V}^{-1}})$, which involves finding the spectral radius of  $(F{{V}^{-1}})$. The calculation yields:
\vspace{-0.5em} % 调整此值来控制公式与前后文本的间距
\begin{align}
\label{R}
  & {{R}_{0}}=(A\sigma ({{a}_{2}}+{{z}_{2}}))/(k\theta ({{m}_{1}}+{{m}_{2}}+{{n}_{2}})({{a}_{1}}+{{a}_{2}}+{{z}_{1}}+{{z}_{2}})) \\ 
   &\quad\quad +(A\sigma ({{a}_{1}}+{{z}_{1}}))/(k\theta ({{m}_{1}}+{{n}_{1}})({{a}_{1}}+{{a}_{2}}+{{z}_{1}}+{{z}_{2}}))\nonumber\
\end{align}
  \vspace{-2em}
\label{eq:my_equation20} % 标签名可以自定义

The propagation threshold ${{R}_{0}}$ determines the speed and scale of public opinion spread; the smaller ${{R}_{0}}$ is, the more favorable it is for controlling public opinion \citep{zhang_multiple_2024}.

\textbf{Proposition 1.}  When ${{R}_{0}}\le 1$, the zero-spread equilibrium point $(\frac{A}{\theta },0,0,0)$ is an asymptotically stable equilibrium point within System Q. When $B(t)$ and $I(t)$ are both 0, there is no public opinion propagation in the system. At this point,  $(\frac{A}{\theta },0,0,0)$ serves as a zero-spread equilibrium point of the system.

\textbf{Proposition 2.}  When ${{R}_{0}}>1$, $(S^*,B^*,{{I}_{1}}^*,{{I}_{2}}^*)$ is a non-zero spread equilibrium point within System Q. Where $S^*=k({{m}_{1}}+{{n}_{1}})\left( ({{n}_{2}}+1)({{a}_{1}}+{{a}_{2}}+1)-({{z}_{2}}+{{a}_{2}}){{m}_{2}} \right)/\sigma \left( ({{n}_{2}}+1)({{z}_{_{1}}}+{{a}_{1}})+({{m}_{1}}+{{n}_{1}})({{z}_{2}}+{{a}_{2}}) \right)$; $B^*=-\theta ({{n}_{2}}+1)S^*/(({{z}_{1}}+{{a}_{1}})({{n}_{2}}+1)+({{z}_{2}}+{{a}_{2}})({{m}_{1}}+{{n}_{2}}))$; ${{I}_{1}}^*=({{z}_{1}}+{{a}_{1}})B^*/k({{m}_{1}}+{{n}_{1}})$; ${{I}_{2}}^*=({{z}_{2}}+{{a}_{2}})B^*/k({{n}_{2}}+1)$.

\section{Situational simulation}
\subsection{Four-party evolutionary game phase}
\subsubsection{Parameter settings}
\label{sec:Parameter} 
The behavior decisions of game subjects are influenced by a variety of factors. To better illustrate the simulation results, this study uses MATLAB R2016b for simulation. This section selects the optimal ideal equilibrium point K (1,1,1,1) as the initial condition for numerical simulation analysis. The study further investigates how varying punishment intensities by the government, targeting different subjects (the perpetrator and the online media), dynamically affect the evolution trajectory of the system.  To eliminate the impact of initial strategy probabilities on system evolution, the initial values of  $x$, $y$, $z$ and $m$ are all set to 0.5.

Due to the complexity of cyber violence, directly obtaining relevant real-world data is more difficult. Therefore, when setting experimental parameters in the model, we primarily adhere to three primary principles: (1) grounding them in actual case data, (2) referencing existing literature, and (3) ensuring consistency with the mathematical logic relationships between parameter. It is worth noting that the model primarily reflects the relationships between variables and does not impact the research conclusions.

To incorporate real-world factors, we reviewed China’s legal and policy documents regarding cyber violence. According to the investigation, China has not yet enacted a specific anti-cyberviolence law and is still in the draft consultation stage. In China, the punishment for cyber violence offenses primarily relies on existing laws, including the \textit{Criminal Law of the People’s Republic of China}, the \textit{Public Security Administration Punishment Law}, and the\textit{ Cybersecurity Law of the People’s Republic of China}. Specifically, according to Article 42, Item (2) of the \textit{Public Security Administration Punishment Law}, those who openly insult others or fabricate facts to defame others, if the circumstances are serious, will face detention of not less than five days and not more than ten days and may also be fined up to 500 RMB\footnote{Public Security Administration Punishment Law of the People’s Republic of China, Article 42, Item (2). (2005). The State Council of the People’s Republic of China. Retrieved from \url{https://www.gov.cn/gongbao/content/2005/content_77704.htm}}.  For participants in cyber violence who act out of ignorance and blindly follow others, their behavior can be classified as other forms of provocation and disturbance. If the circumstances are serious, they may be subject to detention for ten to fifteen days and a fine of up to 1,000 RMB\footnote{Public Security Administration Punishment Law of the People’s Republic of China, Article 26, Item (4). (2005). The State Council of the People’s Republic of China. Retrieved from \url {https://www.gov.cn/gongbao/content/2005/content_77704.htm}}.  In addition, the \textit{Cybersecurity Law of the People’s Republic of China} stresses the network security protection obligations of network operators and imposes a fine of 10,000 RMB or more than 100,000 RMB on those who fail to fulfill their obligations and refuse to make corrections or cause harmful consequences\footnote{Cybersecurity Law of the People’s Republic of China, Article 59. (2016). The State Council of the People’s Republic of China. Retrieved from \url{
https://www.gov.cn/xinwen/2016-11/07/content_5129723.htm}}.  Nevertheless, since the quantitative punishment standards for cyber violence offenses have not yet been clearly defined, and due to the police’s consideration of protecting citizens’ privacy, some data has not been made public, making it difficult to obtain punishment data. Therefore, based on the aforementioned legal provisions and available public case databases, this study sets the initial value  as ${{g}_{1}}=1000$ (assuming it includes both criminal and administrative penalties), ${{g}_{2}}=10000$, $\phi =0.2$, and $\beta =0.4$.

Due to the anonymity of the perpetrator, accurately quantifying the cost of their behavior presents significant challenges. However, it can be inferred that the cost of the perpetrator’s actions is relatively low, given the low cost associated with engaging in cyber violence \citep{marciano_cyberbullying_2020}. Therefore, ${{C}_{p}}$ is set to 10. In contrast, the victim face difficulties in collecting evidence and high costs associated with defending their rights. In contrast, the victim face challenges in evidence collection and bear high costs in defending their rights. These high costs are a significant factor causing many victims to abandon their efforts to defend their rights in practice. Therefore, it can be deduced that ${{C}_{v1}}>{{C}_{p}}$, so we set ${{C}_{v1}}$=1000. The initial values for other parameters are set as follows:  ${{R}_{v1}}=50$, ${{C}_{v2}}=980$, ${{H}_{1}}=15$, ${{H}_{2}}=16$, ${{H}_{3}}=20$, $\delta =5$, ${{L}_{P}}=20$, ${{R}_{P}}=8$, ${{C}_{m1}}=4000$, ${{R}_{m1}}=1500$, ${{R}_{m2}}=2000$, ${{C}_{m2}}=800$, $P=300$, $W=100$, ${{C}_{g1}}=5000$, ${{C}_{g2}}=1000$, ${{F}_{1}}=2000$, ${{F}_{2}}=2500$, ${{F}_{3}}=1000$.

To advance the governance of cyber violence towards a collaborative governance model, the government needs to adopt a “strong regulation” strategy. Punishing the wrongdoings of the perpetrator and the online media by the government is a necessary condition for controlling cyber violence. Next, we will analyze the impact of the government’s punishment severity on the evolutionary outcomes of the system.

\subsubsection{Effect of key variables on the strategy evolution of game subjects}
(1) Effect of punishment severity ($\phi$) on the strategy evolution of game subjects

\begin{figure}[http]
    \centering
    \includegraphics[width=0.5\linewidth]{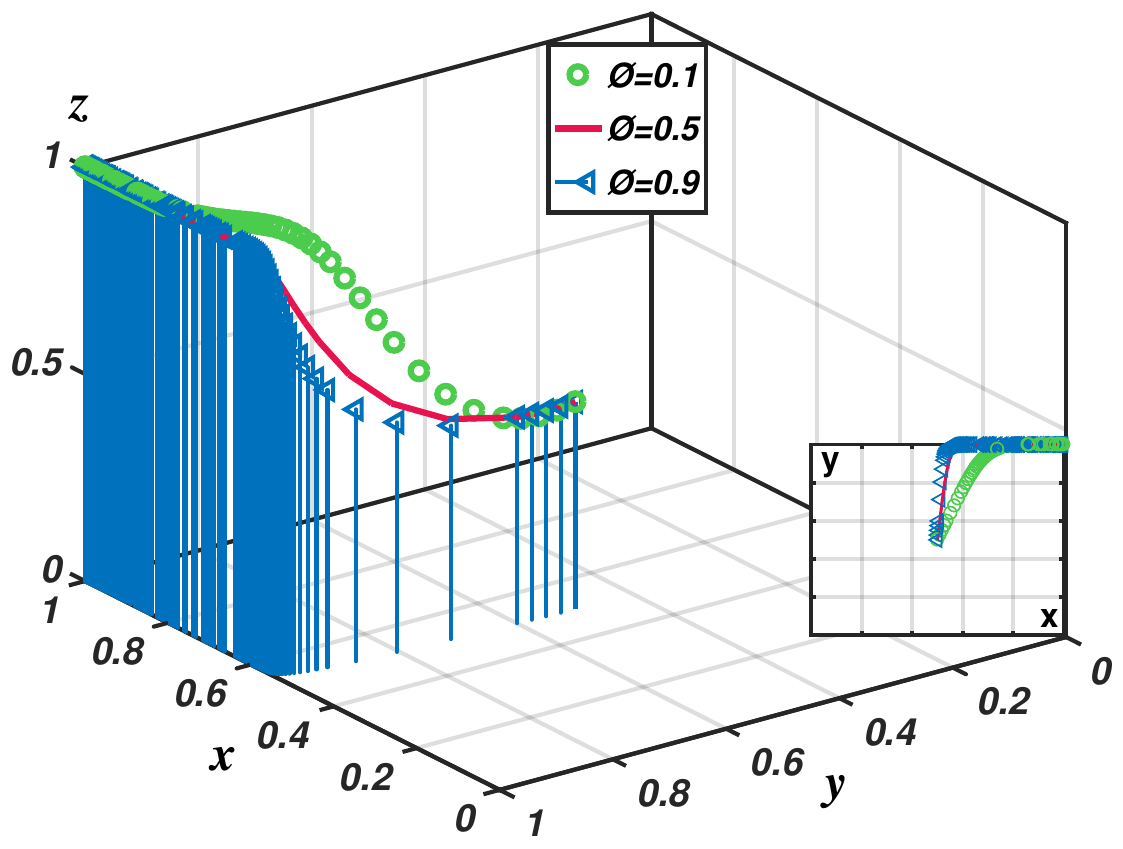}
    \caption{{Effect of punishment severity ($\phi$) on the strategy evolution of game subjects.}}
    \label{fig:Effect of-punishment}
\end{figure}

As shown in \hyperref[fig:Effect of-punishment]{Fig.~\ref*{fig:Effect of-punishment}}, when the government adopts a “strong regulation” strategy to punish the perpetrator’s behavior, the probability that the perpetrator will adopt the “stop attacking” strategy increases as the punishment severity rises. This trend is consistent with the theoretical analysis. As the probability of the government implementing a “strong regulation” strategy increase, the effect of the punishment severity $\phi$ under this strong regulation on the system’s evolutionary outcomes becomes more significant. As the government’s punishment severity towards the perpetrator increases, the probability of the victim taking the “action” strategy decreases, with the victim beginning to rely on the government’s regulation and protection. In this context, the reduced probability of the victim taking “action” and the online media adopting a “correct guidance” strategy implies that the government will need to invest in educational development and proactive guidance efforts.

(2) Effect of punishment severity ($\beta$) on the strategy evolution of game subjects

\begin{figure}[http]
    \centering
    \includegraphics[width=0.8\linewidth]{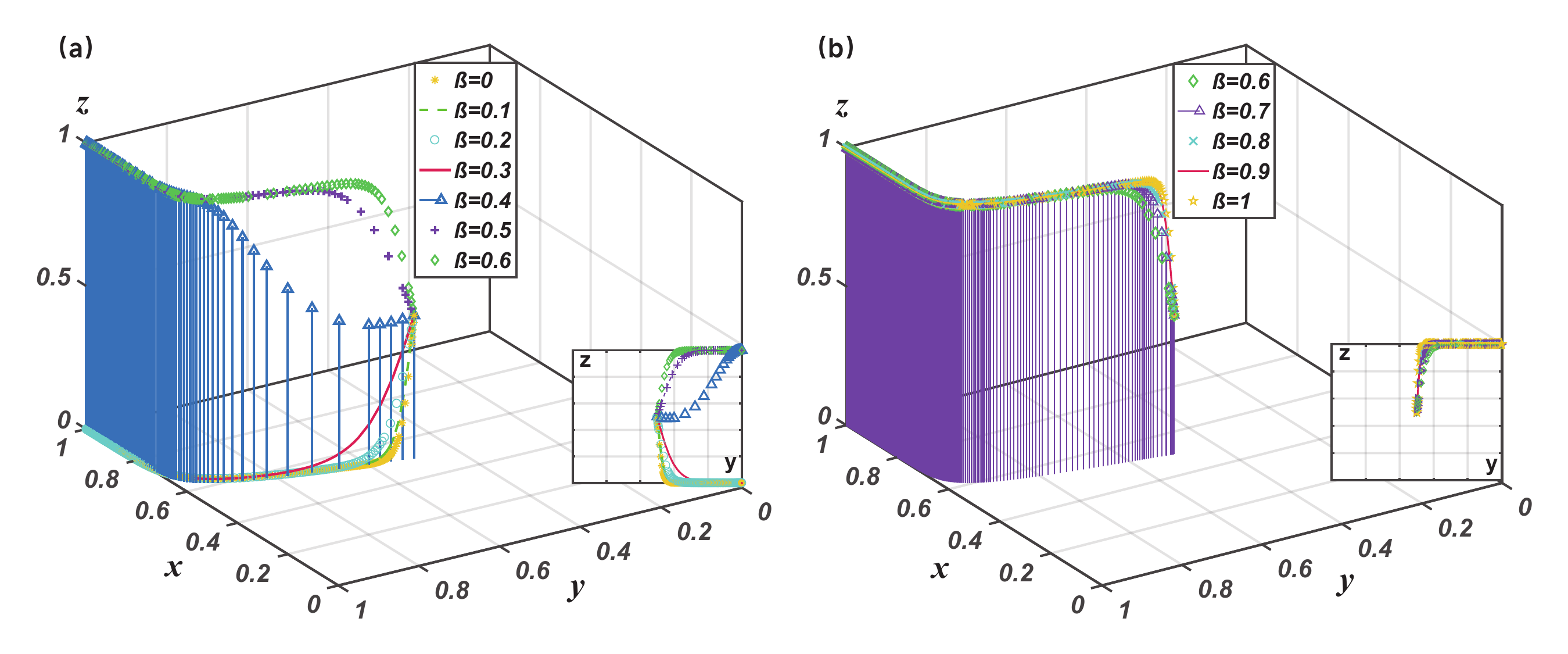}
     \caption{{Effect of punishment severity ($\beta$) on the strategy evolution of game subjects.}}
    \label{fig:Effect-of}
\end{figure}

As shown in \hyperref[fig:Effect-of]{Fig.~\ref*{fig:Effect-of}}, under the influence of strong government regulation, the online media tend to adopt a “correct guidance” strategy, a trend consistent with the theoretical analysis. Moreover, when the government imposes penalties on online media, the deterrent effect gradually increases the likelihood that the perpetrator will adopt a “stop attacking” strategy. As shown in \hyperref[fig:Effect-of]{Fig.~\ref*{fig:Effect-of}}(a), when the punishment severity is at a relatively low level, specifically with a punishment severity $\beta \in [0,0.3]$, the condition $(-{{C}_{m1}}+{{R}_{m1}}-{{R}_{m2}}+P+{{C}_{m2}}+\beta {{g}_{2}})<0$ holds, making G (1,1,0,1) a stable local equilibrium point. This indicates that mild penalties are ineffective in promoting the online media to adopt a “correct guidance” strategy. Because the online media continue to pursue exorbitant profits of and persist in 'misguidance', the system converges only to equilibrium point G. Therefore, increasing the severity of penalties is necessary. When the punishment severity $\beta =0.4$, K (1,1,1,1) becomes a stable local equilibrium point. As the  $\beta$ crosses the threshold of 0.4 and continues to rise to 0.6, the system’s evolution towards the stable strategy significantly accelerates. Nevertheless, as shown in \hyperref[fig:Effect-of]{Fig.~\ref*{fig:Effect-of}}(b), when $\beta $ increases from 0.6 to 0.7, it is evident that the system’s evolution towards this stable strategy gradually slows down. When the $\beta $ further increases, i.e., $\beta \in [0.7,1]$, this accelerating trend significantly weakens, indicating that the marginal effect of increased penalties begins to diminish. Therefore, we can infer that the punishment severity $\beta $ within the range of [0.4, 0.7] can achieve the optimal stable state.  

\subsection{Public opinion dissemination phase}

\subsubsection{parameter initialization settings}
This section employs NetLogo software for simulation, with the target network set as a scale-free network comprising 1,000 nodes. In the small-world network, different states of netizens are represented by “human figures”: pink represents state $S$, purple represents state $B$, blue represents state ${{I}_{1}}$, red represents state ${{I}_{2}}$, and green represents $R$. When $t=0$, the network is set with $S$-number=800 \citep{yu_modeling_2021}, $B$-number=100 \citep{yu_modeling_2021}, ${{I}_{1}}$-number=50 \citep{jiang_dynamic_2020}, ${{I}_{2}}$-number=50 \citep{jiang_dynamic_2020}, and $R$=0 \citep{geng_online_2023}. The range for all other parameters is set between [0,1], with specific settings detailed in \hyperref[Description-of]{Table~\ref{Description-of}}.

\begin{table}[H]%% placement specifier
%% Use tabular environment to tag the tabular data.
%% https://en.wikibooks.org/wiki/LaTeX/Tables#The_tabular_environment
\centering%% For centre alignment of tabular.
\fontsize{9}{11}\selectfont % Set the font size to 9pt
\renewcommand{\arraystretch}{1} % Adjust row height
\caption{\hspace{0.5em}Description of model parameters.}
\label{Description-of}
\begin{tabular}{p{1.3cm} p{4cm} p{1.3cm} p{9cm}} 
%% Tabular cells are separated by &
  \toprule 
  Parameters & Descriptions & Initial value & Source \\ % Header row
  \hline
  $A$ & The rate that new users enter the networks & 1.0 & \rev{See Hypothesis 3 in \hyperref[sec:Basic assumptions of the public opinion]{Section 3.2.2}} \\
  $\sigma $ & State transition probability from $S$ to $B$ & 0.2 & \rev{Referring to the parameter settings for simulating network information dissemination in Section 3 Experiment of \citet{zhang_sei3r_2022}.} \\
  ${{z}_{1}}$ & Probability of correct guidance by the online media & 0.5 & \rev{The parameter baseline was set at 0.5, following \citet{jaynes_probability_2003} maximum entropy principle to avoid subjective bias without prior information, and aligning with the modeling paradigm of symmetric probability baselines in experiment as employed by \citet{camerer_behavioral_2003}.}\\
  ${{z}_{2}}$ & Probability of misguidance by the online media & 0.5 & \rev{The same as above.}\\
    ${{m}_{1}}$ & Probability of strong government regulation & 0.5 & \rev{The same as above.}\\
  ${{m}_{2}}$ & Probability of weak government regulation & 0.5 & \rev{The same as above.}\\
  ${{a}_{1}}$ &  Probability of positive emotional identification under the correct guidance of the online media & 0.5 & \rev{Refer to Section 4 Data Fitting in \citet{yinModellingDynamicEmotional2021}, Section 4 Numerical Simulations in \citet{yanEmotioninformationSpreadingModel2024}, and Section 4 Numerical Simulation of the Model in \citet{zhang_multiple_2024}.}\\
  ${{a}_{2}}$ & Probability of negative emotional identification under the misguidance of the online media & 0.5 & \rev{The same as above.}\\
  $k$  & Government credibility & 0.5 & \rev{We assumed that the initial value $k$=0.5, reflecting an equilibrium state where public trust is neither polarized nor lacking \citep{Dalton}, providing a relatively neutral and reasonable starting basis for the subsequent evolution of the model.} \\
  ${{n}_{1}}$  & Probability of positive emotional identification under the strong regulation of the government & 0.9 & \rev{Based on the consideration of strong government regulation, we assumed that the initial value of ${{n}_{1}}$ is 0.9.}\\
  ${{n}_{2}}$& Probability of positive emotional identification under the weak regulation of the government & 0.1 & \rev{Refer to Section 3 SPNR: The Proposed Model in \citet{bao_precise_2015}.} \\
  $\theta $ & State transition probability from $S$ to $R$ & 0.1 & \rev{Given the relatively low probability of netizens spontaneously changing from a susceptible state to an immune state, and we reasonably set the initial value of the $\theta $ to 0.1.} \\
  \bottomrule % 底部粗线 
\end{tabular}
\end{table}

\subsubsection{Simulation experiment}
\begin{figure}[H]
    \centering
    \includegraphics[width=0.7\linewidth]{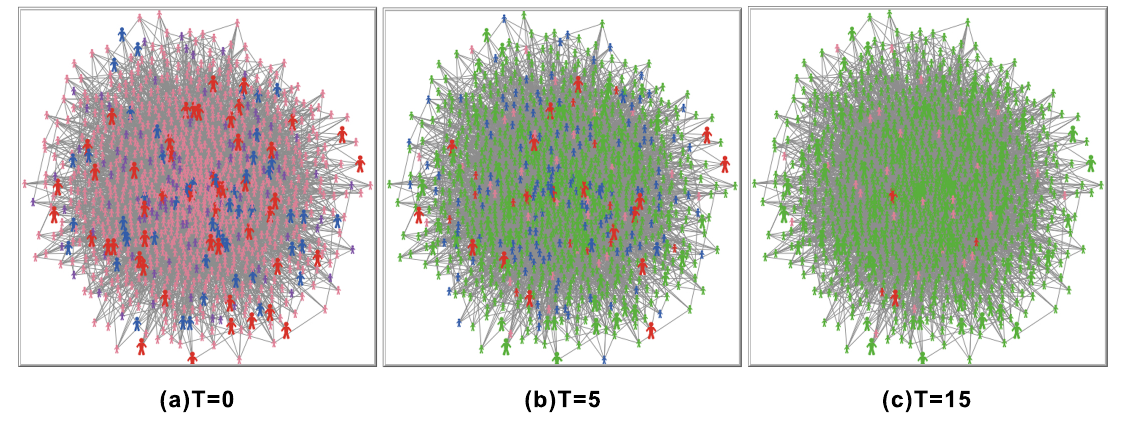}
     \caption{{The diffusion process of cyber violence public opinion among online user groups.}}
    \label{fig:The diffusion process of cyber violence public opinion among online user groups.}
\end{figure}

Based on the given initial conditions, a simulation was conducted to depict the basic structure of the cyber violence public opinion propagation model. The results represent the average of 50 iterations, as shown in \hyperref[fig:The diffusion process of cyber violence public opinion among online user groups.]{Fig.~\ref*{fig:The diffusion process of cyber violence public opinion among online user groups.}}.  

\hyperref[fig:The diffusion process of cyber violence public opinion among online user groups.]{Fig.~\ref*{fig:The diffusion process of cyber violence public opinion among online user groups.}} shows that online users continuously change their states over time as the government and the online media implement their intervention strategies. Ultimately, the disseminators of pro-cyberviolence will gradually withdraw from the system. In the specific diffusion process of cyber violence public opinion, different behavioral intervention strategies can influence both the speed and the state of netizens’ transitions. Next, we further discuss the impact of different behavioral strategy probabilities and combinations employed by the online media and the government on the propagation of public opinion.\rev{As the control group, the baseline values of ${{z}_{1}}$ and ${{m}_{1}}$ were both set to 0.5, i.e., all parameter values for the control group in \hyperref[Description model-of]{Table~\ref{Description model-of}} are derived from \hyperref[Description-of]{Table~\ref{Description-of}}. The parameter values for the other experimental groups are shown in \hyperref[Description model-of]{Table~\ref{Description model-of}}. Based on the control group’s baseline conditions, this study further explores the impact of two dimensions:}\rev{Firstly, the impact of subject strategy probability on cyber violence public opinion propagation. We set ${{z}_{1}}$ to 0.1 and 0.9, with corresponding values of ${{z}_{2}}$ being 0.9 and 0.1, which formed experimental group 1 and experimental group 2, respectively. Similarly, when ${{m}_{1}}$ was set to 0.1 and 0.9, the corresponding values of ${{m}_{2}}$ were set to 0.9 and 0.1, which formed experimental group 3 and experimental group 4. Secondly, the impact of subject intervention strategies on cyber violence public opinion propagation. When ${{z}_{1}}$ was set to 1 and ${{m}_{1}}$ to 0, with corresponding values of ${{z}_{2}}$ set to 0 and ${{m}_{2}}$ to 1, this formed experimental group 5 (i.e., only online media positive intervention ). When ${{z}_{1}}$ was set to 0 and ${{m}_{1}}$ to 1, with corresponding values of ${{z}_{2}}$ set to 1 and ${{m}_{2}}$ to 0, this formed experimental group 6 (i.e., only government positive intervention). When both ${{z}_{1}}$ and ${{m}_{1}}$ were set to 1, with corresponding values of ${{z}_{2}}$ set to 0 and ${{m}_{2}}$ to 0, this formed experimental group 7 (i.e., a collaborative positive intervention by both online media and the government). Furthermore, $\theta $ was adjusted to 0.3 and k was adjusted to 0.6 on the basis of experimental group 7, forming experimental group 8.}

\begin{table}[H]%% placement specifier
%% Use tabular environment to tag the tabular data.
%% https://en.wikibooks.org/wiki/LaTeX/Tables#The_tabular_environment
\centering%% For centre alignment of tabular.
\fontsize{9}{11}\selectfont % Set the font size to 9pt
\renewcommand{\arraystretch}{1} % Adjust row height
\caption{\hspace{0.5em}Description of model parameters.}
\label{Description model-of}
\begin{tabular}{p{3.5cm} p{0.7cm} p{0.7cm} p{0.7cm} p{0.7cm}p{0.7cm}p{0.7cm}p{0.7cm}p{0.7cm}p{0.7cm}p{0.7cm}p{0.7cm}}  
%% Tabular cells are separated by &
  \toprule 
  \diagbox[width=3cm]{Group}{Parameter} & $\sigma $ & ${{z}_{1}}$ & ${{z}_{2}}$ & ${{m}_{1}}$ & ${{m}_{2}}$ & ${{a}_{1}}$ & ${{a}_{2}}$ & $k$ & ${{n}_{1}}$ & ${{n}_{2}}$ & $\theta$ \\ % Header row
  \hline
  Control Group & 0.2	& 0.5 &	0.5	& 0.5 & 0.5 & 0.5 & 0.5 & 0.5 & 0.9 & 0.1 & 0.1 \\
  Experimental Group 1 & 0.2	& 0.1 &	0.9 & 0.5 & 0.5 & 0.5 &	0.5 & 0.5 &	0.9 & 0.1 &	0.1 \\
  Experimental Group 2 & 0.2	& 0.9 &	0.1 & 0.5 & 0.5 & 0.5 &	0.5 & 0.5 &	0.9 & 0.1 &	0.1\\
  Experimental Group 3 & 0.2 & 0.5 & 0.5 & 0.1 & 0.9 & 0.5 & 0.5 & 0.5 & 0.9 & 0.1 & 0.1\\
  Experimental Group 4 & 0.2 & 0.5 & 0.5 & 0.9 & 0.1 & 0.5 & 0.5 & 0.5 & 0.9 & 0.1 & 0.1\\
  Experimental Group 5 & 0.2	& 1 & 0 & 0 & 1 & 0.5 & 0.5 & 0.5 & 0.9 & 0.1 & 0.1 \\
  Experimental Group 6 & 0.2	& 0 & 1 & 1 & 0 & 0.5 & 0.5 & 0.5 & 0.9 & 0.1 & 0.1 \\
  Experimental Group 7 & 0.2	& 1 & 0 & 1 & 0 & 0.5 & 0.5 & 0.5 & 0.9 & 0.1 & 0.1 \\
  Experimental Group 8& 0.2	& 1 & 0 & 1 & 0 & 0.5 & 0.5 & 0.6 & 0.9 & 0.1 & 0.3 \\
  \bottomrule % 底部粗线 
\end{tabular}
\end{table}

(1)	Impact of subject strategy probability on cyber violence public opinion propagation

\begin{figure}[http]
    \centering
    \includegraphics[width=0.7\linewidth]{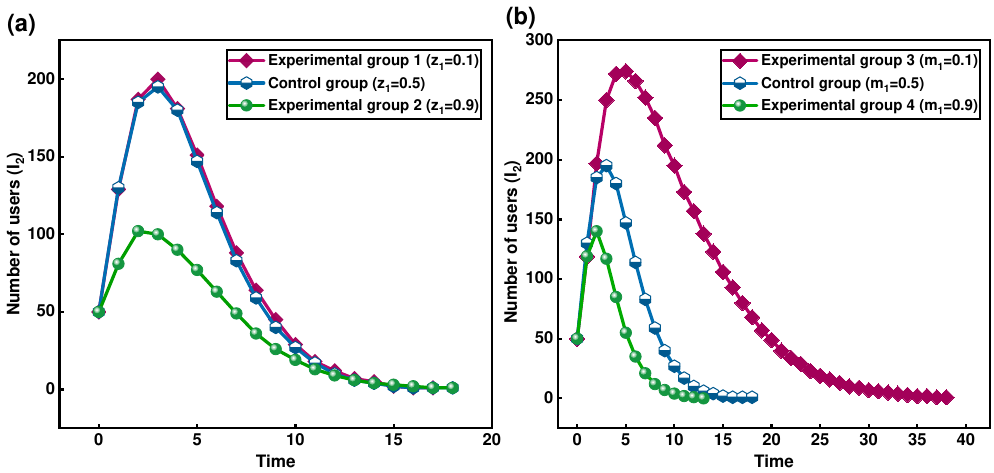}
     \caption{{Impact of initial probabilities of the online media and government on the density of disseminators of pro-cyberviolence.}}
    \label{fig:Impact of initial probabilities of the online media and government on the density of disseminators of pro-cyberviolence.}
\end{figure}

We explore the impact of initial strategy probabilities of the online media and government on the propagation of cyber violence public opinion. The simulation results for online media are shown in \hyperref[fig:Impact of initial probabilities of the online media and government on the density of disseminators of pro-cyberviolence.]{Fig.~\ref*{fig:Impact of initial probabilities of the online media and government on the density of disseminators of pro-cyberviolence.}}(a), while those for government are shown in \hyperref[fig:Impact of initial probabilities of the online media and government on the density of disseminators of pro-cyberviolence.]{Fig.~\ref*{fig:Impact of initial probabilities of the online media and government on the density of disseminators of pro-cyberviolence.}}(b). The peak number of propagator nodes supporting cyber violence  $\max \{{{I}_{2}}(t)\}$ can be used to measure the maximum severity of public opinion propagation. The  $\max \{{{I}_{2}}(t)\}$ represents the highest density of the population supporting cyber violence.
 
Based on \hyperref[R]{Eq.(20)}, when ${{z}_{1}}$ is 0.1, 0.5, and 0.9, the propagation thresholds ${{R}_{0}}$ are 3.40, 3.25, and 3.09, respectively. The propagation threshold gradually decreases, indicating an improvement in the effectiveness of controlling cyber violence public opinion. As shown in \hyperref[fig:Impact of initial probabilities of the online media and government on the density of disseminators of pro-cyberviolence.]{Fig.~\ref*{fig:Impact of initial probabilities of the online media and government on the density of disseminators of pro-cyberviolence.}}(a), as the proportion of ${{z}_{1}}$ increases, the number of individuals transitioning from bystanders to disseminators of pro-cyberviolence decreases. The $\max \{{{I}_{2}}(t)\}$ also decreases, leading to a reduced spread of public opinion. It is important to note that while solely increasing the proportion of the online media’s “correct guidance” strategy can reduce the range of the impact of cyber violence public opinion and decrease the time for infection nodes ${{I}_{2}}(t)$ to fall to zero, it does not significantly alter the timing of the value of $\max \{{{I}_{2}}(t)\}$ for disseminators of pro-cyberviolence. Cyber violence public opinion has continued to propagate and spread within a certain range in cyberspace and the victim will endure greater public pressure.

 As shown in \hyperref[fig:Impact of initial probabilities of the online media and government on the density of disseminators of pro-cyberviolence.]{Fig.~\ref*{fig:Impact of initial probabilities of the online media and government on the density of disseminators of pro-cyberviolence.}}(b), with the increasing proportion of ${{m}_{1}}$, propagators of cyber violence gradually transition to immune individuals. The  $\max \{{{I}_{2}}(t)\}$ significantly decreases, and the timing of  $\max \{{{I}_{2}}(t)\}$ appears significantly earlier, with a notable reduction in the duration of public opinion propagation. This indicates that the proportion of the government’s “strong regulation” strategy has a significant effect on controlling the spread of public opinion. 

(2)	Impact of subject intervention strategies on cyber violence public opinion propagation

\begin{figure}[http]
    \centering
    \includegraphics[width=0.7\linewidth]{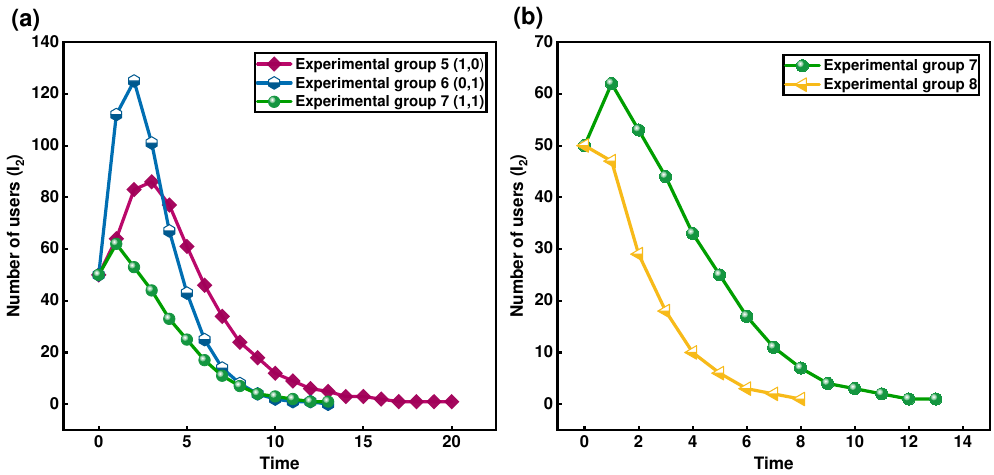}
     \caption{{(a) Impact of different intervention strategies on the spread of public opinion; (b) Impact of government credibility and immunity rate on the spread of public opinion.}}
    \label{fig: (a) Impact of different intervention strategies on the spread of public opinion; (b) Impact of government credibility and immunity rate on the spread of public opinion.}
\end{figure}

To thoroughly analyze the positive effects of collaborative governance between the online media and government in controlling public opinion dissemination, we established three experimental scenarios: (a) Experimental group 5 represents positive intervention by the online media alone, with a combination strategy of (1,0); (b) Experimental group 6 represents positive intervention by the government alone, with a combination strategy of (0,1);  (c) Experimental group 7 represents collaborative positive intervention by both the online media and government, with a combination strategy of (1,1). The specific simulation results are shown in \hyperref[fig: (a) Impact of different intervention strategies on the spread of public opinion; (b) Impact of government credibility and immunity rate on the spread of public opinion.]{Fig.~\ref*{fig: (a) Impact of different intervention strategies on the spread of public opinion; (b) Impact of government credibility and immunity rate on the spread of public opinion.}}(a).

As shown in \hyperref[fig: (a) Impact of different intervention strategies on the spread of public opinion; (b) Impact of government credibility and immunity rate on the spread of public opinion.]{Fig.~\ref*{fig: (a) Impact of different intervention strategies on the spread of public opinion; (b) Impact of government credibility and immunity rate on the spread of public opinion.}}(a), the (1,0) strategy is ineffective in controlling public opinion in the short term, indicating that correct guidance by the online media alone cannot effectively curb the spread of public opinion. When the online media adopts a “misguidance” strategy and lacks effective government regulation, netizens develop incorrect emotional perceptions of cyber violence events, which can cause bystanders to gradually become disseminators of pro-cyberviolence, thereby prolonging the dissemination of cyberviolence-related public opinion. Therefore, a single positive intervention strategy by either the online media or the government is insufficient to effectively curb the spread of cyber violence. In fact, it may even exacerbate its dissemination and impact, leading to more widespread negative effects. Based on the comparison of propagation thresholds, it can be concluded that the suppression of public opinion is most effective when both the online media and government intervene positively, followed by government intervention alone while relying solely on the online media intervention yields the least effective results.

To quickly control public opinion spread, enhancing government credibility and netizen resilience is crucial. In Experimental Group 8, we set $k=0.6$ and $\theta =0.3 $, comparing it with Experimental Group 7. The results (\hyperref[fig: (a) Impact of different intervention strategies on the spread of public opinion; (b) Impact of government credibility and immunity rate on the spread of public opinion.]{Fig.~\ref*{fig: (a) Impact of different intervention strategies on the spread of public opinion; (b) Impact of government credibility and immunity rate on the spread of public opinion.}}(b)) show that increasing government credibility and immunity rate reduces the propagation threshold (${{R}_{0}}=0.69<1 $), significantly decreasing pro-cyberviolence disseminators and shortening the dissemination duration, thus rapidly controlling the public opinion crisis.

\section{Case study}
To further validate the operability of collaborative governance mechanism for cyber violence in real-world cyber violence events, this study uses the cases of  “Chongqing man (codename: Fat Cat) jumping into the river, which triggered cyber violence against his girlfriend, \rev{known by the pseudonym Tan Zhu”} and “Zhang’s hiring of ‘online troll groups’ to cyber violence against others” (due to space constraints, see Appendix G. for this case) as research examples. The selection criteria are as follows: (1) Typicality: This study selected representative cyber violence incidents from the past two years in China that garnered significant public attention and discussion. (2) Diverse manifestations of the events (See Appendix A.): The diverse forms of manifestation will strongly support the argument for the model’s universality and applicability. (3) Involvement of multiple interest groups and formation of online public opinion. (4) Availability and verifiability of case data.

\begin{figure}[H]
    \centering
    \includegraphics[width=0.8\linewidth]{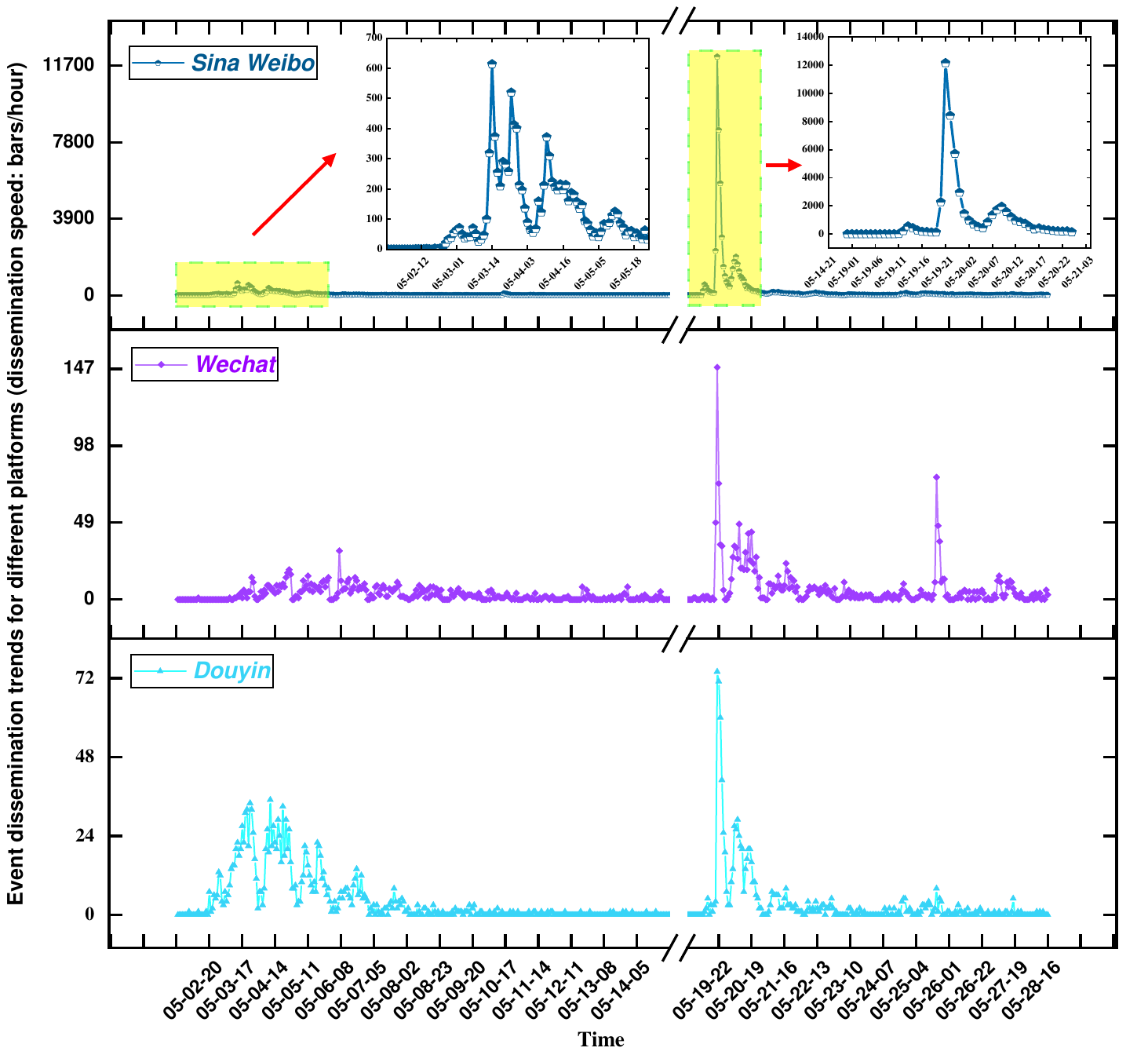}
     \caption{{Event propagation for different platforms.}}
    \label{fig:Event-propagation}
\end{figure}

First, we utilized data from the Zhiwei Shijian platform (China’s leading tool for monitoring and analyzing public opinion, \url{
https://ef.zhiweidata.com/}) to create a trend chart illustrating the dissemination of the Fat Cat river-jumping incident across different platforms, as shown in \hyperref[fig:Event-propagation]{Fig.~\ref*{fig:Event-propagation}}. Specifically, users can retrieve relevant incidents by registering and logging into the Zhiwei Shijian platform, selecting the enterprise advanced version service, and entering the keyword “Fat Cat” in the search bar of the event database. By clicking on the “Dissemination Trend” option, users can access detailed dissemination trend data across various platforms. Next, the incident was divided into five stages based on the lifecycle theory and the data, i.e., formation, outbreak, continuation, secondary outbreak, and subsidence stage. Finally, simulations were conducted for each stage using the collaborative governance mechanism for cyber violence above. 
The incident sparked widespread cyber violence and triggered several derivative discussions, raising significant challenges for cyberspace governance. We have mapped the dissemination data across three online platforms, covering the periods from 00:00 on May 2 to 24:00 on May 14 and from 00:00 on May 15 to 16:00 on May 28, as detailed in \hyperref[fig:Event-propagation]{Fig.~\ref*{fig:Event-propagation}} Compared to the WeChat and Douyin platforms, Sina Weibo exhibits the highest dissemination popularity. Additionally, several Weibo topics related to the incident rapidly gained traction. For example, the Weibo topics \#\textit{21-year-old Chongqing Man Jumps into River}\# (with 350 million views), \#\textit{Tan Zhu Response}\# (with 730 million views), \#\textit{Fat Cat and Tan Zhu}\# (with 220 million views), \#\textit{Fat Cat’s Sister Speaks Out}\# (with 150 million views), and \#\textit{Police Report on Economic Transactions between Fat Cat and Tan}\# (with 1.4 billion views). The data mentioned in the above topics can be found directly in the Related Topics screen of Sina Weibo\rev{(https://www.weibo.com/)}. The high reading volumes on Sina Weibo made it a hotspot, prompting the selection of Weibo platform data for analysis in this section.

The Fat Cat incident triggered two rounds of public opinion storms. From \hyperref[fig:Event-propagation]{Fig.~\ref*{fig:Event-propagation}},
it can be observed that the dissemination rate between May 2 and May 14, peaked at 644 messages per hour on May 3 at 2:00 PM, lasting 13 days. During this period, Fat Cat’s girlfriend, Tan, faced significant online harassment. After the release of the Chongqing police report, a new wave of discussion emerged, lasting 9 days and 16 hours, with a peak dissemination rate of 12,545 messages per hour on May 19 at 9:00 PM. The event had a long fermentation period and intense public attention, surpassing the severity of other major social events at the time.

\subsection{Formation stage: April 11, 2024 – May 1, 2024}

\begin{figure}[http]
    \centering
    \includegraphics[width=0.8\linewidth]{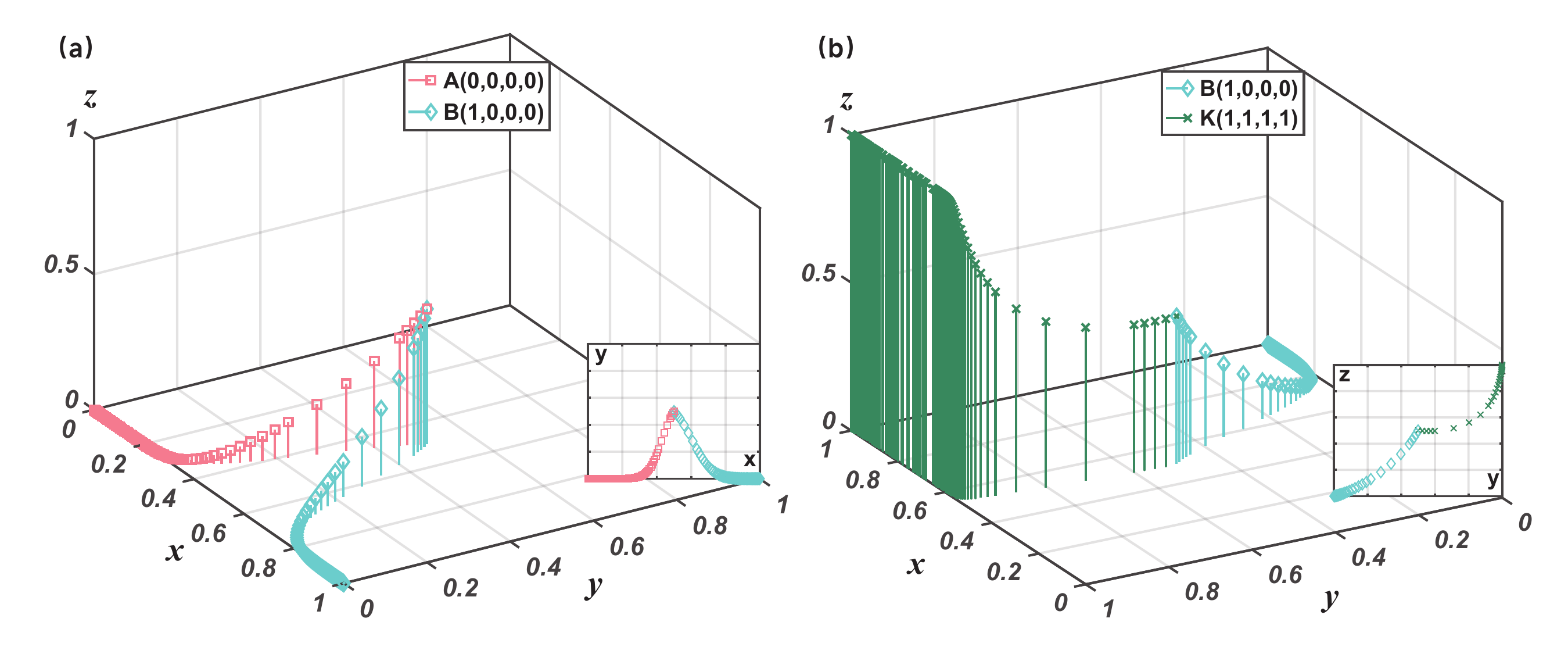}
     \caption{{(a) Evolutionary paths of equilibrium points A and B; (b) Evolutionary paths of equilibrium points B and K.}}
    \label{fig:Evolutionary paths-of}
\end{figure}

On April 11, 2024, Fat Cat, a popular streamer in the Esports boosting community, died by jumping into a river in Chongqing, China. Two weeks later, the perpetrator (Liu,\rev{a pseudonym}, Fat Cat’s sister) posted a video on Douyin platform, claiming the suicide was due to an emotional dispute, which initially drew little public attention.  Subsequently, the victim, Tan (Fat Cat’s girlfriend) became a target of cyber violence. Tan subsequently reported the case to public security, reflecting the transition from A (0,0,0,0) to B (1,0,0,0) in Scenario 3 (see Appendix E.). The parameter settings are consistent with the mathematical logic between equilibrium points A and B, as illustrated in \hyperref[fig:Evolutionary paths-of]{Fig.~\ref*{fig:Evolutionary paths-of}}(a), \rev{which is consistent with the robustness discussed above.} The victim, pressured by personal and external factors, tended to adopt the “action” strategy. The perpetrator's attacks and online media's misguidance intensified cyber violence, while the government’s initial “weak regulation” strategy failed to contain the incident and escalated social governance costs and risks.

\subsection{Outbreak stage: May 2–3, 2024} 

On May 2, 2024, news that “21-year-old ‘Fat Cat,’ was coerced into giving over 500,000 RMB by his online girlfriend under the pretext of love, and later jumped into the Yangtze River after she broke up with him” topped Sina Weibo’s trending list. The perpetrator’s claim of seeking justice and the victim’s denial further fueled the discourse. Some online media outlets, disregarding journalistic ethics, employed sensationalist and inappropriate tactics to attract attention, exacerbating the escalation of cyber violence. The spread of misleading opinions not only damaged online media credibility but also incited attacks on the victim before the police investigation concluded. The victim persisted to report the case, while the perpetrator’s attacks and online media’s misguidance continued, causing severe societal harm, and the strategic choices of all parties stabilized at equilibrium point B (1,0,0,0), consistent with the formation stage. Online media’s misguidance can harm victims through privacy breaches and psychological stress while encouraging the perpetrator’s attacks, disrupting cyberspace order. Though it may offer short-term gains, this strategy risks reputational damage and government sanctions, making it unsustainable.

With a total reading volume of 1.45 billion for Weibo topics like \#\textit{21-year-old Chongqing Man Jumps into River }\#, \# \textit{Tan Zhu Response} \#, etc., the susceptible population is assumed to be 1,450. Additionally, to simplify the analysis, we assume that the number of disseminators for both anti-cyberviolence and pro-cyberviolence is 50.\rev{At the same time, during the outbreak phase, we conducted data collection, processing, sentiment analysis, and validation. The specific steps are as follows:}

\rev{(1) Data crawling: As indicated by the above analysis, online media continue to adopt misguidance strategies. Therefore, under the premise of adhering to Weibo's terms of service and regulations, we used Python tools to retrieve comment text data from the API endpoint of Sina Weibo. This data was collected from 00:00 on May 2, 2024, to 24:00 on May 3, 2024, based on the misleading viewpoints published by five popular online media accounts (@9t9lBlihvP, @i8xDVthQlK, @sd1GQPcB89, @UhJFttu0q0 and @Zl1vNCOUHR. These are all anonymous. The specific anonymization steps and code can be found in item (d) in the “Note” of Appendix A and item (a) of Appendix I.), totaling 2,782 comments (the screenshot of the crawler program code can be found in Appendix H.).}

\rev{(2) Anonymization process: We adhere to ethical guidelines for anonymization. For all the final data obtained in this study, we performed hash anonymization (using the SHA-256 irreversible hash algorithm) on the personal usernames. The specific process and code can be found in Appendix H, item (b).}

\rev{(3) Data cleaning: The initial dataset of 2,782 entries was cleaned using Python, which included removing duplicate entries, deleting advertisements, filtering out irrelevant information related to the research topic, and eliminating unnecessary special characters such as \%, \#, etc. As a result, 2,078 valid comments were retained.}

\rev{(4) Sentiment analysis and validation:
We manually annotated the sentiment of the final valid dataset, classifying the sentiments as either positive or negative. The annotations were carried out by two PhD students with a background in information management using the Doccano annotation platform (see the Appendix H.). The inter-annotator agreement was evaluated using the Brennan-Prediger coefficient, which reached 0.8536 (more than 0.6), and Cohen's Kappa coefficient, which was 0.8047 (more than 0.8). Both evaluation metrics exceeded 0.8, indicating a high level of consistency in the annotations. In cases of annotation discrepancies, a domain expert was invited to make the final adjudication. The final results showed that negative sentiment accounted for 75.07\% of the dataset. Based on this result, we reasonably assumed the negative sentiment parameter $a_2 = 0.75$.} 

The cyber violence escalated into a public opinion storm, assuming ${{R}_{0}}>1$.\rev{Based on the aforementioned analysis of the evolutionary path, the system stabilizes at the equilibrium point B(1,0,0,0). Therefore, we set $z_1 = 0$ and $m_1 = 0$, with corresponding values $z_2 = 1$ and $m_2 = 1$ (since $z_1 + z_2 = 1$, $m_1 + m_2 = 1$).}  Based on the characteristics of the outbreak stage, we assumed $\sigma =0.3$, $\theta =0.15$, and ${{n}_{2}}=0.2$. The other parameters remain consistent with the control group. At this point,\rev{we calculate that ${{R}_{0}}=3.58$}. The simulation evolution is shown in \hyperref[fig:Number of ${{I}_{2}}$ in different stages.]{Fig.~\ref*{fig:Number of ${{I}_{2}}$ in different stages.}}(a).

\begin{figure}[http]
    \centering
    \includegraphics[width=0.9\linewidth]{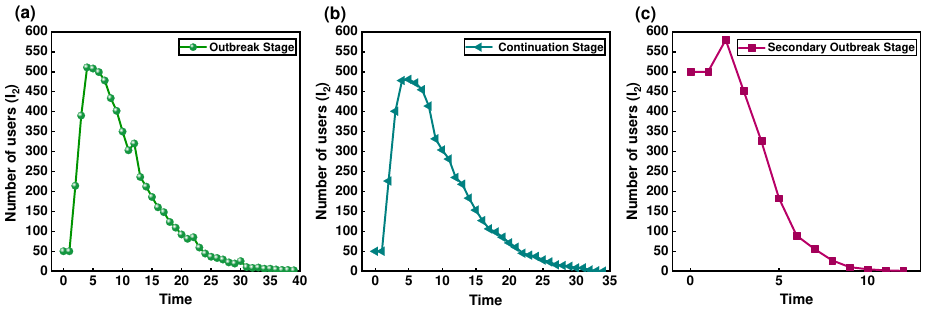}
    \caption{{Number of ${{I}_{2}}$ in different stages.}}
    \label{fig:Number of ${{I}_{2}}$ in different stages.}
\end{figure}
As shown in \hyperref[fig:Number of ${{I}_{2}}$ in different stages.]{Fig.~\ref*{fig:Number of ${{I}_{2}}$ in different stages.}}(a), the population density of pro-cyberviolence disseminators initially accelerates, peaks, and then gradually declines, with fluctuating trends during the decline. This reflects a complex, long duration of dissemination due to the lack of proactive intervention by the government and online media. Starting on May 3, an online memorial activity was initiated, but improper handling by merchants led to secondary public opinion crises over issues like “empty packages” and “plain water.”  Topics like
 \#\textit{Final Handling of Food Delivery}\# and \#\textit{Food Waste}\# worsened the public opinion crisis, challenging the collaborative governance of cyber violence.
 
\subsection{Continuation stage: May 4–5, 2024}  

Public opinion remained in a continuation stage, with a propagation threshold lower than that of the outbreak stage. As the number of susceptible individuals converting to immunes increased, $\theta $ was set to 0.2, and ${{R}_{0}}$ was calculated to be 2.68.  The other parameters remained the same as in the outbreak stage. The simulation evolution is shown in \hyperref[fig:Number of ${{I}_{2}}$ in different stages.]{Fig.~\ref*{fig:Number of ${{I}_{2}}$ in different stages.}}(b). The peak density of disseminators of pro-cyberviolence during this stage is lower than that during the outbreak stage, yet the peak remains at a relatively high level, posing a threat to social governance.

\subsection{Secondary outbreak stage: May 19–20, 2024}  

\rev{Around 20:00} on May 19, 2024, the Public Security Bureau’s announcement explicitly stated that the victim was not involved in criminal fraud, and public opinion surged once again. The perpetrator disclosed a large amount of information related to the victim’s privacy and employed methods such as hiring content writers and purchasing online traffic to mislead netizens. After the police report, Chinese government departments including @CCTV News, @People’s Daily, and @The Beijing News, actively spread information via online platforms. The perpetrator and online media outlets that disseminated misleading viewpoints began to fear government sanctions. The perpetrator tended to adopt a “stop attacking” strategy, while the online media leaned towards a “correct guidance” strategy. This stage aligns with the evolution from the equilibrium point B (1,0,0,0) in scenario 3 to the equilibrium point K (1,1,1,1) in scenario 4. The detailed evolution process is illustrated in \hyperref[fig:Evolutionary paths-of]{Fig.~\ref*{fig:Evolutionary paths-of}}(b),\rev{which is consistent with the robustness discussed above.}

Based on the previously mentioned Weibo topic \# \textit{Police Report on Economic Transactions between Fat Cat and Tan} \#, which garnered 1.4 billion views, the susceptible population is assumed to be 1,400. Based on the observation of the data, there are many more disseminators in this stage than in the continuation stage, and we assume that the number of disseminators of anti-cyberviolence and pro-cyberviolence is 500.\rev{Based on the time of the announcement issued by the Public Security Bureau and the measures taken by online media for correct guidance, during the secondary outbreak phase, we used Python to scrape 2,443 comments with correct guidance viewpoints from five popular online media platforms (@TELnO06mIl, @8vr7EL7C6M, @cxfxMiJTOZ, @Ep4bCg3ZyE, and @Paoyya5kyM. These are all anonymous.) between 20:00 on May 19, 2024, and 20:00 on May 20, 2024. The data processing and sentiment classification process were consistent with those applied during the initial outbreak stage. In the end, 2,159 valid text entries were obtained. Similarly, the data processing, sentiment analysis, and verification steps were consistent with those of the outbreak phase. Among them, the Brennan-Prediger coefficient was 0.8759, and Cohen’s Kappa coefficient was 0.8397. The final result also showed that positive sentiment text accounted for 25.71\%. Based on this result, we reasonably assumed that $a_1 = 0.26$.} Despite correct guidance from the five online media outlets, the reversal of the event made the perpetrator shift targets, sparking new negative sentiments. This suggests that relying solely on online media's guidance is insufficient to address complex cyber violence, aligning with prior theoretical analysis.

%As shown in \hyperref[fig:Event-propagation]{Fig.~\ref*{fig:Event-propagation}}, 

{\hypersetup{linkcolor=purple}As shown in \hyperref[fig:Event-propagation]{Fig.~\ref*{fig:Event-propagation}}}, the peak transmission speed during this stage is higher than that of the first outbreak phase, indicating a greater heat of public opinion. Therefore, we assumed that the $R_0$ in this stage is larger than that in the first outbreak stage.\rev{Based on the aforementioned analysis of the evolutionary path, the system stabilizes at the equilibrium point K(1,1,1,1). Therefore, we set $z_1 = 1$ and $m_1 = 1$, with corresponding values $z_2 = 0$ and $m_2 = 0$ (since $z_1 + z_2 = 1$, $m_1 + m_2 = 1$). Considering the characteristics of the secondary outbreak stage, such as a larger value of $R_0$, we assumed $\sigma = 0.4$.} The other parameters remain consistent with the control group.\rev{At this point, we calculate that \( R_0 = 5.10 > 3.58 \) (first outbreak stage).} The simulation evolution is shown in \hyperref[fig:Number of ${{I}_{2}}$ in different stages.]{Fig.~\ref*{fig:Number of ${{I}_{2}}$ in different stages.}}(c). The density of pro-cyberviolence disseminators briefly peaks before sharply declining to near zero. Although the density of disseminators supporting cyber violence initially remained high, the government’s credibility, strong regulations, and online media guidance ultimately led perpetrators to cease attacks, while netizens shifted to posting positive viewpoints, rapidly controlling public opinion. {\hypersetup{linkcolor=purple}This is also supported by \hyperref[fig:Event-propagation]{Fig.~\ref*{fig:Event-propagation}}}, which shows that the transmission speed at 24:00 on May 20 (142 messages/hour) was lower than at 24:00 on May 3 (213 messages/hour).
%This is also supported by \hyperref[fig:Event-propagation]{Fig.~\ref*{fig:Event-propagation}},

\subsection{Subsidence stage: May 6–18, 2024, and May 21–28, 2024}   

From May 6 to May 18, 2024, the Chinese government investigated and collected evidence. Despite no further public opinion escalation, the lack of effective intervention measures disrupted social order. From May 21 to May 28, 2024, government departments focused on guiding rational discussions via official accounts.

It is important to note that, despite challenges in data quantification and collection, this study has made every effort to gather relevant data to ensure that the simulation model's parameters accurately reflect real-world conditions. This enhances the accuracy and reliability of results when simulating cyber violence stages and scenarios. By integrating case background information, the study aligns simulation outcomes with specific contexts and stakeholder behaviors, thereby developing a comprehensive interpretative framework. This improves the fusion of simulation experiments with case studies, enhancing the research's scientific rigor and practical relevance.

\subsection{Simulation validation}

\begin{figure}[htbp]
    \centering
    \includegraphics[width=0.6\linewidth]{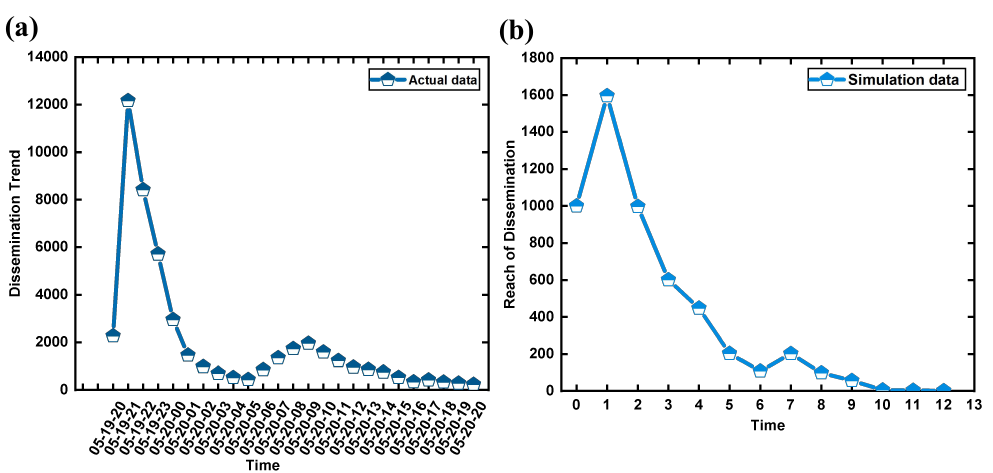}
    \caption{%
        Comparison of simulation outcomes with real case data. \\
        \footnotesize{Note: Reach of Dissemination in (b) represents the sum of the number of disseminators $I_1$ and $I_2$.}
    }
    \label{fig:Comparison of simulation outcomes with real case data.}
\end{figure}

\rev{By analyzing the public opinion evolution characteristics of the “Fat Cat” incident, it can be observed that a typical pattern of the government and online media collaborative intervention emerged during the secondary outbreak stage (${{z}_{1}}=1$, ${{m}_{1}}=1$), which is an objective fact that can be confirmed. This also provides a crucial observational window for model validation. We compared the actual diffusion curve during this critical phase with the diffusion trajectory output by the simulation model, where the simulation data represents the average result of 50 independent runs. As shown in \hyperref[fig:Comparison of simulation outcomes with real case data.]{Fig.~\ref*{fig:Comparison of simulation outcomes with real case data.}}, the simulation results are generally consistent with the real-world diffusion curve in terms of the growth or decay trends of public opinion evolution. This aligns with the widely accepted academic understanding that model validation requires demonstrating that the average results of the model correlate with real-world outcomes. The validation results indicate that our model accurately captures real-world behavior to some extent.}

\vspace{-0.5cm}
\section{Discussions}
\subsection{Principal findings}
In the evolutionary game phase, an important finding in our study was that moderate punishment of misguiding online media by the government is optimal (\hyperref[fig:Effect-of]{Fig.~\ref*{fig:Effect-of}}), consistent with the inverted U-shaped curve theory \citep{menetre_temporal_2023} of punishment effects, which suggests that both excessively strong and weak punishments reduce effectiveness, with the optimal being of moderate intensity. This is inconsistent with the prior study that advocated harsher penalties for illegal media, a view supported by \citet{xie_influence_2021}. But they did not fully examine the complex relationship between punishment severity and its effect. As a adaptive subject, the online media adjust strategies based on government measures. Moderate punishment helps balance rule compliance and profit-seeking, enabling sustainable social governance. This study provides new evidence on punishment mechanisms for misguiding media. Additionally, 
we found that the correct guidance of the online media helped encourage victims to take “action,” revealing the collaborative role of objective media reporting in governing cyber violence.

In terms of government punishment towards the perpetrator, our study revealed that under the government's “strong regulation,” the likelihood of the perpetrator choosing to cease the attack increased as the severity of punishment rose.
This finding is consistent with previous research \citep{masmoudi_unveiling_2025}. Additionally, when the government imposes punishment on the online media, the probability of the perpetrator adopting the “stop attacking” strategy gradually increases, which may reflect the deterrent effect of government actions. This deterrence not only directly impacts the perpetrator but also potentially alters their expectations regarding risks and consequences, prompting them to adjust their behavior. This revealed the potential of government to normalize social behavior in cyber violence governance through enhanced punitive measures.

Another finding was that the victim tends to adopt the “action” strategy under proactive government guidance. However, as the government increases punishment against the perpetrator, the victim is less likely to take “action” and tends to rely more on government regulation and protection. This may be due to the high cost of rights protection and the psychological and social pressure victims face \citep{zych_systematic_2015}. As a result, the victim may prioritize maximizing their benefits by hoping the government imposes harsher penalties on the perpetrator to reduce the burden of personal rights protection. The decreased likelihood of the victim taking “action” suggests that the government needs to invest more in education, guidance, and psychological support. To our knowledge, current research lacks an examination of victims' dependence on government protection. This gap may be related to privacy concerns regarding victims’ identities and the ethical review challenges of data anonymization, as well as cultural differences. In existing policy frameworks, victims are often not treated as independent entities. For instance, in the European Union's Digital Services Act (hereinafter referred to as the DSA) \citep{Digital}, victims were seen as part of “user satisfaction,” which focused on platform efficiency while neglecting their psychological needs and reliance on government support. However, in multi-agent governance, victims can also be active participants \citep{marcum_editorial_2024,caneppele_observing_2025}, such as reporting perpetrators or sharing their experiences to promote social co-governance. Future research should focus on the multiple roles and behavioral strategies of victims, such as how to design refined policies to meet their needs and offer greater opportunities for governance participation, thus facilitating the healthy development of the cyber violence governance system.

In the public opinion dissemination phase, this study demonstrated through simulations that, under controlled variables, the proportion of online media adopting a “correct guidance” strategy negatively correlated with the density of disseminators of pro-cyberviolence, consistent with the normative focus in social influence theory \citep{stok_focus_2019}. Specifically, increasing the probability of “correct guidance” strengthens societal expectations of correct behavior, thus reducing the density of disseminators of pro-cyberviolence. This is because individuals are often influenced by social norms when making behavioral choices. However, while this strategy shortened the time to reduce pro-cyberviolence to zero, it did not significantly delay the time to peak for disseminators of pro-cyberviolence. This revealed a key dilemma in cyber violence governance: relying solely on positive guidance from online media could not effectively curb the outbreak phase \citep{wang_modeling_2024}. Failure to address this may increased the chilling effect \citep{zhu_speaking_2021, chin_comparative_2022} on netizens, leading to desensitization or indifference toward cyber violence. Over time, this have destabilized cyberspace. Therefore, future interventions should be adjusted \citep{marcum_examining_2019}. For example, this study suggests increasing the probability of the government’s “strong regulation” strategy concurrently to achieve the goals of good governance and sustainable development \citep{achuthan_cyberbullying_2023}.

Furthermore, increasing the probability of both “correct guidance” by the online media and “strong regulation” by the government can significantly shortened the time to peak for disseminators of pro-cyberviolence. This finding validated the applicability of the institutional complementarity theory \citep{aoki_toward_2001} in cyberspace: the coupling of formal regulation and informal norms \citep{chang2023}, which, through the “information correction-deterrence enhancement” dual-loop mechanism, overcame the dilemma of the failure of positive guidance. Moreover, the analysis in \hyperref[fig: (a) Impact of different intervention strategies on the spread of public opinion; (b) Impact of government credibility and immunity rate on the spread of public opinion.]{Fig.~\ref*{fig: (a) Impact of different intervention strategies on the spread of public opinion; (b) Impact of government credibility and immunity rate on the spread of public opinion.}}(a) showed the ranking of the effectiveness of suppressing public opinion on cyber violence: collaborative positive intervention $>$ government strong regulation alone $>$ online media correct guidance alone. This finding provides mathematical support for numerous studies on collaborative governance \citep{rodriguez-rodriguez_how_2020,ng_effectiveness_2022}.

\subsection{Adaptation of the cyber violence governance framework}

\rev{Due to differences in the legal, social, and technological environments across regions, the governance framework for cyber violence should be adapted to local contexts. Our cyber violence governance framework possesses a certain degree of general applicability. For example, in countries such as Singapore and Vietnam, governments often intervene directly in content moderation and offender punishment through legislative mandates. The logic of our framework offers direct reference value for such governance contexts. The European Union's DSA emphasizes platform responsibility and content transparency, with a governance logic that follows a layered model of “legislative empowerment, platform self regulation, and regulatory fallback.” Taking the DSA as an example, we propose the following adaptive adjustments to our cyber violence governance framework, including but not limited to:
}

\rev{Evolutionary game phase:}

\rev{(1) Adaptive adjustments for the online media}

\rev{Firstly, the European Union's DSA provides a clear delineation of responsibilities for online platforms, offering an important basis for analyzing the role of online platforms in cyber violence governance. The DSA requires online platforms to establish risk management systems, actively monitor and remove illegal content, and regularly report their content moderation measures to regulatory authorities. These provisions strengthen the key role of online platforms in cyber violence governance. Based on this, if the focus of the research is on the influence of online platforms, the agent of online media can be adjusted to that of online platforms accordingly.}

\rev{Secondly, the DSA explicitly states that online platforms have the responsibility to actively monitor and remove illegal content. Therefore, the strategic options available to online platforms can be adjusted to {strict moderation, lenient moderation}. Strict moderation refers to online platforms adopting stringent content review and blocking mechanisms, actively monitoring and removing harmful content. Lenient moderation refers to online platforms maintaining openness, implementing a more relaxed content moderation system.}

\rev{Thirdly, online platforms are required to regularly report their content moderation and transparency measures to regulatory authorities (see Article 24 in the DSA). While this entails certain costs, standardized reporting can also enhance platform transparency and compliance, potentially leading to greater benefits. Therefore, separate cost and benefit parameters can be established for this aspect.}

\rev{Fourthly, the DSA mandates that online platforms provide user complaint-handling systems (see I, Item (58)), which incurs additional operational costs for the platform, such as setting up and maintaining complaint channels and the labor costs of handling complaints. However, effective user protection measures can increase user trust in the platform, which, in turn, may influence the willingness of victims to take action and the behavioral strategies of perpetrators. Separate cost and benefit parameters can also be set here to reflect the actual impact of this measure.}

\rev{Fifthly, the DSA stipulates that online platforms violating regulations will be subject to fines (see Article 74 in the DSA). Introducing a fine parameter can reflect the economic risks that platforms may face. This not only increases the cost for perpetrators to continue their attacks but also pressures platforms to strengthen their oversight of cyber violence content under compliance demands, thereby forcing perpetrators to adjust their behavioral strategies.}

\rev{(2) Adaptive adjustments for the government
In the context of the European Union, the government's role may focus more on enforcing various laws and regulations through platform oversight mechanisms, representing an indirect form of regulation. For example, the DSA requires online platforms to establish risk management systems, such as increasing algorithmic transparency (see Items (65) of I and Article 24 in the DSA). Therefore, the government's regulatory costs should include the costs associated with auditing the compliance of online platforms.}

\rev{(3) Adaptive adjustments for the perpetrator}

\rev{Due to the strict regulation of illegal content under the DSA, the cost for the perpetrator to continue their attacks will increase. Therefore, we can assume that under the environment of strict content moderation on online platforms, the perpetrator is more likely to stop their attacks. For instance, the probability of “stop attacking" could be adjusted as $y \in [0.5, 1]$.}

\rev{Accordingly, in this study, the entity intervening in the public opinion dissemination phase has changed from the online media to online platforms.}

\rev{With the above adaptive adjustments, our cyber violence governance framework can better align with the legal and social context of the European Union. This analysis not only helps to understand regional differences in cyber violence diffusion but also offers valuable insights for policymakers to make targeted adjustments based on their own legal systems and social environments. Future research can collect more data to further validate these assumptions and parameters, enhancing the scalability and applicability of the governance framework. Although the situations in other countries vary widely, similar adaptive adjustments can still be made.}

\subsection{Theoretical implications}
The theoretical contribution of this study lies in providing a novel perspective for understanding the multi-agent strategic interactions and public opinion dissemination mechanisms in cyber violence. 

Firstly, Previous research has primarily focused on the adversarial relationship between regulatory authorities and perpetrators \citep{glance2024cyberviolence}, while overlooking the impact of the online media and victim  on the behavior of other actors. By incorporating evolutionary game theory, this study offers an in-depth examination of the dynamic interactions among multiple stakeholders in cyber violence, thereby extending the application of game theory. This study proposes a four-party evolutionary game framework involving the victim, the perpetrator, the online media, and the government, which more profoundly reveals the behavioral strategies and expectations of multiple agents and more comprehensively reflects the real-world context of cyber violence.

Secondly, conventional studies have typically analyzed punitive measures from a qualitative perspective and as a singular approach {\citep{Zhu2024}}. In contrast, this study innovatively introduces a dual-dimensional framework that considers both the targets and intensity of punishment. This framework provides a scientific basis for regulatory authorities to formulate differentiated sanctioning strategies, effectively addressing the limitations of the single-dimensional punitive paradigm.

Thirdly, in the field of cyber violence public opinion dissemination, although existing studies have identified the role and influence of bystanders in cyber violence, most have adopted a static perspective, failing to capture the dynamic evolution of bystander attitudes and their heterogeneity. In reality, the macro-level evolution of group polarization stems from complex interactions among individuals \citep{dongUsingSimulation2022}. In response, this study innovatively incorporates the bystander role and develops an optimized cyber violence public opinion propagation model. Additionally, drawing on recent research findings, the model integrates emotional factors of bystanders \citep{tao2025, luSentiment2024} to examine the micro-level dynamics of their transition from neutrality to either supporting or opposing cyber violence. Furthermore, we incorporate game-theoretic strategies to analyze how public opinions evolve under different intervention strategies implemented by the government and online media. This study not only provides policymakers with a quantitative reference for intervention strategies but also actively responds to the theoretical advancements in collaborative governance 
\citep{ghenai2025,rodriguez-rodriguez_how_2020}. Our study offers a new pathway for addressing the challenges posed by online group polarization. 

In conclusion, this study framework integrates government, societal, and individual emotional and behavioral factors, effectively bridging the gap between macro and micro level analyses in cyber violence research. By drawing on well-established theories such as evolutionary game theory, propagation dynamics, and emotional contagion, this study  not only deepens the understanding of complex diffusion mechanisms in cyber violence but also provides novel insights into collaborative governance mechanisms, which help to promote the transformation of cyber violence governance from single-perspective approaches to integrated collaborative models.

\subsection{Practical implications}
Our study has broad practical implications, encompassing victims, perpetrators, bystanders, online media, and policymakers. 

Firstly, for victims, regulatory authorities can leverage collaborative governance and adaptive mechanisms to dynamically adjust the strategic parameters of stakeholders, thereby maintaining a stable and balanced online environment. Additionally, policymakers should encourage rights protection and continuously optimize relief mechanisms by leveraging intelligent platforms to streamline legal aid processes, ensuring that victims receive timely and effective support. 

Secondly, regarding perpetrators and bystanders, regulatory authorities should quantify penalties for perpetrators while employing sentiment analysis techniques for real-time monitoring and intervention. Intervention strategies should be adapted to different stages of public opinion development. For instance, during the emergence phase, media guidance protocols can be activated. During the outbreak phase, measures to mitigate group polarization should be deployed, and multi-platform emergency coordination should be initiated to ensure effective intervention by regulatory authorities and online media.  

Thirdly, for the online media, policymakers can enhance proactive guidance by establishing a structured public opinion guidance matrix. Additionally, agenda-setting strategies can be employed to redirect public attention, increasing the likelihood of transitioning susceptible individuals to resistant ones. This approach is theoretically consistent with the findings of \cite{penny2021} on attention-based interventions. 

Finally, policymakers can establish dedicated anti-cyberviolence information platforms (e.g., an official website for a cybersecurity knowledge base) to systematically integrate educational resources such as legal interpretations, case analyses, and prevention guidelines, thereby enhancing public awareness and protective capabilities. Policymakers can also build an information exchange platform for collaborative governance between the government and society, and further improve the reporting mechanism to ensure a healthy online environment.

\subsection{Limitations and future works}
In future research, Some limitations should be considered and optimized in future research. Firstly, the case data used in the current study have certain limitations. Due to the difficulty in obtaining real-world data, we collected as much data as possible and integrated case information for the analysis.\rev{The model of cyber violence public opinion propagation cannot fully capture the interactions among the victim, perpetrator, government, and online media during the dissemination process. Future research will seek breakthroughs through advancements in big data technology and the integration of interdisciplinary theories, with a} focus on the collection and analysis of large-scale data and the development of big data models that provide rich information to guide subsequent research directions. This involves thorough exploration and analysis of all elements of cyber violence incidents, as well as technological innovations for detection. Secondly, the generalizability of the results should be exercised with caution. Our research did not fully consider cultural differences, so future studies should place more emphasis on validation and comparison across different cultural contexts. Furthermore, interdisciplinary collaboration and international exchanges will be crucial to advancing cyber violence research.\rev{Finally, Although bystanders have been incorporated into the SB$\mathrm{I}_1$$\mathrm{I}_2$R model, their specific emotional and behavioral heterogeneity has not been fully considered. Future work will introduce more fine-grained emotional and behavioral differentiation to more effectively capture the complexity of real-world scenarios.}

\section{Conclusions}

This study combines evolutionary game theory and diffusion dynamics, considering both macro and micro factors, and proposes a new two-stage, multi-scenario collaborative governance mechanism for cyber violence. In the evolutionary game phase, we found that cyber violence governance is influenced by the joint of the victim, perpetrator, online media, and government. The evolutionarily stable strategy of a subject depends not only on its own interests but also on the strategies of other players. Secondly, the government's punishment strategy should be adjusted according to different subjects. In the public opinion dissemination phase, the effectiveness of interventions follows this order: collaborative positive intervention $>$ government strong regulation alone $>$ online media correct guidance alone. Finally, through a multi-stage and multi-scenario simulation analysis of real cases, we further enhance the practical application value of this study. This study offers new theoretical insights and practical guidance for understanding the strategic choices of stakeholders in cyber violence, predicting public opinion dynamics, and establishing a robust, long-term collaborative governance mechanism for cyber violence.

\singlespacing % 
\bibliography{elsarticle-template-harv}
\bibliographystyle{elsarticle-harv}

\end{document}